\documentclass{amsart}
\usepackage[latin1]{inputenc}
\usepackage[dvips]{graphics}
\newtheorem{proposition}{Proposition}
\newtheorem{theorem}{Theorem}
\newtheorem{lemma}{Lemma}

\usepackage{bbm,changebar,ulem}

\def\ind{\mathbbm{1}}

\def\C{{\mathbb C}}
\def\N{{\mathbb N}}
\def\R{{\mathbb R}}

\def\P{{\mathbb P}}
\def\E{{\mathbb E}}

\def\eps{\varepsilon}
\def\mean{\mathbb{E}}

\def\I{\mathbb{I}}

\def\etal{{et al.}}

\newcommand\marg[1]{ }

\title[Perturbation analysis of an $M/M/1$ queue]{Perturbation analysis of an $\mathbf{M/M/1}$ queue in a diffusion random environment}
\author{Christine Fricker}
\address[Christine Fricker]{INRIA Paris --- Rocquencourt,  Domaine de Voluceau, 78153
 Le Chesnay, France}
\email{Christine.Fricker@inria.fr}
\author{Fabrice Guillemin}
\address[Fabrice Guillemin]{France Telecom R\&D, CORE/CPN, 22300 Lannion, France}
\email{Fabrice.Guillemin@orange-ftgroup.com}
\author{Philippe Robert}
\address[Philippe Robert]{INRIA Paris --- Rocquencourt,   Domaine de Voluceau, 78153
 Le Chesnay, France}
\email{Philippe.Robert@inria.fr}
\urladdr{http://www-rocq.inria.fr/\~{}robert}


\begin{document}

\begin{abstract}
We study  in this paper an  $M/M/1$ queue whose server  rate depends upon the  state of an
independent Ornstein-Uhlenbeck diffusion process $(X(t))$ so that its value at time $t$ is
$\mu \phi(X(t))$, where $\phi(x)$ is some bounded function and $\mu>0$. We first establish
the differential  system for the conditional  probability density functions  of the couple
$(L(t),X(t))$ in  the stationary regime,  where $L(t)$ is  the number of customers  in the
system at time  $t$. By assuming that  $\phi(x)$ is defined by $\phi(x)  = 1-\varepsilon (
(x\wedge a/\varepsilon)\vee(-b/\varepsilon))$ for some  positive real numbers $a$, $b$ and
$\varepsilon$, we show that the above differential system has a unique solution under some
condition on $a$  and $b$. We then show  that this solution is close,  in some appropriate
sense, to  the solution to the differential  system obtained when $\phi$  is replaced with
$\Phi(x)=1-\varepsilon  x$ for  sufficiently  small $\varepsilon$.  We  finally perform  a
perturbation analysis of  this latter solution for small $\varepsilon$.  This allows us to
check at the first order the validity of the so-called reduced service rate approximation,
stating  that  everything happens  as  if  the server  rate  were  constant  and equal  to
$\mu(1-\eps\E(X(t)))$.
\end{abstract}

\maketitle

\section{Introduction}
We consider in this paper an $M/M/1$ queue with a  server rate varying in time. We
specifically assume that the server rate  at time $t$  is equal to $\mu \phi(X(t))$ for some
function $\phi$ and some auxiliary process $(X(t))$. Throughout this paper, we shall
assume that the modulating process $(X(t))$ is an Ornstein-Uhlenbeck  process with mean
$m>0$, drift coefficient $\alpha >0$, and diffusion coefficient $\sigma >0$. This process
satisfies It\^o's stochastic equation 
\begin{equation}
\label{itosde}
dX(t) = -\alpha (X(t) -m) dt + \sigma dB(t),
\end{equation}
where $(B(t))$ is a standard Brownian motion.  The stationary distribution of the process $(X(t))$ is a normal distribution with mean  $m$ and variance
$\sigma^2/(2\alpha)$; the associated  probability density function is defined on the whole of $\R$ and is given by  
\begin{equation}
\label{normal}
n(x)\stackrel{\text{def.}}{=}\frac{1}{\sigma}\sqrt{\frac{\alpha}{ \pi}}\exp\left(-\frac{\alpha (x-m)^2}{\sigma^2}\right).
\end{equation}
Throughout this paper,  we shall assume that the Ornstein-Uhlenbeck process is
stationary. 

If $L(t)=j$ denotes the number of customers in the $M/M/1$ queue and $X(t)=x$ at time $t$, then the transitions of the process  $(L(t))$ are given by
$$
j \to \left\{\begin{array}{lll} j+1 & \mbox{with rate} & \lambda , \\ j-1 & \mbox{with rate} & \mu\phi(X(t)) .\end{array}\right.
$$
In the following, we shall assume that the condition $\rho\stackrel{\text{def.}}{=}{\lambda}/{\mu} < \mean(\phi(X(0))) \leq 1$
is satisfied so that it is straightforward to show the existence of a stationary
probability distribution for the Markov process $(X(t), L(t))$; see Meyn and Tweedie~\cite{Meyn:01} for example. 

The  study of  the  above system  is  motivated by  the problem  of  bandwidth sharing  in
telecommunication networks and  the coexistence on the same  transmission links of elastic
traffic,  which adapts  to the  level of  congestion of  the network  by achieving  a fair
sharing of  the available  bandwidth, and unresponsive  traffic, which  consumes bandwidth
without taking care  of other traffic. See for instance  \cite{Massoulie} for a discussion
about bandwidth sharing  in packet networks. The choice  of an Ornstein--Uhlenbeck process
as modulating  process is natural for  several reasons: mathematically this  is a standard
``typical'' diffusion process with an equilibrium distribution and secondly it can be seen
as  a  centered approximation  of  the  number of  jobs  of  an  $M/M/\infty$ queue  (the
unresponsive     traffic),     see     for     example     Borovkov~\cite{Borovkov}     or
Iglehart~\cite{Iglehart} or Chapter~6 of Robert~\cite{Robert:08}.

One of the objectives of this paper is to investigate  the so-called Reduced Service Rate
(RSR) property for which the system would behave as if  the server rate were equal to
the mean value $\mu \mean(\phi(X_0))$. Even though some results can be established for
arbitrary perturbation functions $\phi(x)$, we shall pay special attention in the
following to the case when the function $\phi(x)$ has the form 
\begin{equation}
\label{formphi}
\phi(x) = 1-\varepsilon ((x\wedge (a/\varepsilon))\vee (-b/\varepsilon))
\end{equation}
for some small $0<\varepsilon<1$ and real numbers $0<a<1$ and $b>0$, where we use the
notation $a\vee b = \max(a,b)$ and $a\wedge b = \min(a,b)$. The
choice of the bounded perturbation function is discussed at the end of the paper.

As  it  will  be  seen,  one  of  the important  technical  problems  encountered  in  the
perturbation analysis is the existence of a reasonably smooth density probability function
for the  couple $(X(t),L(t))$  in the stationary  regime. Conditions on  $\varepsilon$ for
ensuring  the existence  and the  uniqueness  of a  density probability  function will  be
established in the following via Hilbertian analysis. \cbstart 
More precisely, let $p_j(x)$ denote the stationary  probability density function that the process $(L(t))$ is in state $j$ and the process $(X(t))$ is in state $x$ and let $P$ denote the vector whose $j$th component is $p_j(x)/\rho^j$. In a fist step, we show that $P$ is solution to an equation of the type
\begin{equation}
\label{eqbase}
\Omega f + V(\phi)  f = 0,
\end{equation}
where $\Omega$ is a selfadjoint second order differential operator on $\mathcal{D}'(\R)^\N$, $\mathcal{D}'(\R)$ denoting the set of distributions in $\R$ . Unfortunately, the operator $V(\phi)$ is not selfadjoint so that Kato's perturbation theory for selfadjoint operators cannot be applied. Nevertheless, we prove that the above equation has a unique non null smooth solution $P \in C^2(\R)^\N$ for sufficiently small $\varepsilon$. In addition, we prove that when replacing $\phi(x)$ with $\Phi(x) = 1-\varepsilon x$, we obtain an  equation of the type
\begin{equation}
\label{eqbasebis}
\Omega f + \varepsilon V(\Phi)  f = 0,
\end{equation}
which has a unique non null smooth solution for sufficiently small $\varepsilon$, $V(\Phi)$ being independent of $\varepsilon$. Denoting this solution by $g$, we prove that $P$ and $g$ are close  to each other for some adequate norm when $\varepsilon$ is small. We then perform a power series expansion in $\varepsilon$  of $g$ and we determine the radius of convergence of this series. By explicitly computing the two first terms of the series, this eventually enables us to prove the validity of the reduced service rate approximation at the first order for the system. 
\cbend

%To compute the stationary distribution of the number of customers in the above $M/M/1$ queue, we establish in a first step the Fokker-Planck equations of the system and we show that
%these equations can be seen as an eigenvalue problem for a self-adjoint operator defined
%in some adequate Hilbert space. This last property is closely related to the
%time-reversibility property satisfied by the $M/M/1$ occupation process and the Ornstein-Uhlenbeck
%process. We then show that when the interaction between the modulating  process and
%the $M/M/1$ queue is weak and depends upon a small parameter $\varepsilon$, the problem of
%computing the generating function of the number of customers in the $M/M/1$ queue can be
%solved by means of a regular perturbation analysis.   When the perturbation function has the form given by Equation~\eqref{formphi}, we  completely compute
%the coefficients of the expansion in power series of $\varepsilon$ of the solution. Also,
%the radius of convergence is determined. By taking into account the first order only, the
%above analysis shows that the RSR approximation pertains. 

The problem  considered in this  paper falls into  the framework of queueing  systems with
time  varying server rate,  which have  been studied  in the  queueing literature  in many
different  situations. In  Núñez-Queija and  Boxma~\cite{Nunez1}, the  authors  consider a
queueing system where  priority is given to some flows driven  by Markov Modulated Poisson
Processes (MMPP) with  finite state spaces and the low priority  flows share the remaining
server capacity according to the  processor sharing discipline.  By assuming that arrivals
are Poisson and service times are  exponentially distributed, the authors solve the system
via    a    matrix    analysis.     Similar    models   have    been    investigated    in
Núñez-Queija~\cite{Nunez3,Nunez2}  by  still  using  the  quasi-birth  and  death  process
associated  with  the system  and  a  matrix analysis.   The  integration  of elastic  and
streaming  flows has been  studied by  Delcoigne \etal~\cite{Proutiere},  where stochastic
bounds for the mean number of active flows have been established.  More recently, priority
queueing systems with  fast dynamics, which can  be described by means of  quasi birth and
death processes, have been studied via a perturbation analysis of a Markov chain by Altman
{et al}~\cite{Altman}. A probabilistic analysis  of these queues with varying service rate
has been  presented in  Antunes \etal~\cite{Antunes:03,Antunes:01}. Our  point of  view in
this paper is completely different since  a functional analysis approach is used to tackle
the perturbation  analysis.  The  key difficulty  for the case  considered in  the present
paper is that the associated Markov chain has an infinite state space.

This paper is organized as follows: In Section~\ref{Fokker}, we establish the basic system
of partial differential equations for the joint probability density functions of the process
$(X(t),L(t))$. We recall in Section~\ref{generator} some basic results on the generators
of the Ornstein-Uhlenbeck process and the occupation process in an $M/M/1$ queue. 
In Section~\ref{existence} it is proved that this system has a unique solution with
convenient regularity properties in an adequate
Hilbert space \cbstart when $\varepsilon$ is sufficiently small. In Section~\ref{perturbation}, \sout{we carry out a perturbation analysis for the
perturbation function $\Phi(x)=1-\varepsilon x$}, we show that when replacing $\phi(x)$ with $\Phi(x) = 1-\varepsilon x$, the corresponding differential system has also a smooth solution in the underlying Hilbert space when $\varepsilon$ is sufficiently small. \cbend We then prove  that the solutions to the
differential systems for $\phi$ defined by Equation~\eqref{formphi} and $\Phi$ are close
 to each other in some appropriate sense. \cbstart By expanding the solution of the second differential system in power series of $\varepsilon$, \cbend we  show that
at the first order the so-called Reduced Service Rate property for the original system holds;
the subsequent terms of the associated expansion are  also expressed. Some concluding remarks
are presented in Section~\ref{conclusion}.

\section{Fundamental differential problem}\label{Fokker}

\subsection{Notation and differential system}

The goal of this section is to establish the fundamental differential system for the
conditional probability density functions $p_j(x)$, $j \geq 0$,  in the stationary regime,
where $p_j(x)$ is the probability that the process $(L(t))$ is in state $j$ knowing that
the Ornstein-Uhlenbeck process $(X(t))$ is in state $x$. As long as we do not have proved
regularity results, these functions have to be considered in the sense of  distributions,
i.e., for all $j \geq 0$, $p_j(x) \in \mathcal{D}'(\R)$, where $ \mathcal{D}'(\R)$ is the
set of distributions in $\R$. The distributions $p_j(x)$, $j \geq 0$, are formally defined
as follows: for every infinitely differentiable function with compact support $\varphi(x)$
(denoted, for short, $\varphi \in C^\infty_0(\R)$) 
$$
\int_\R p_j(x) n(x) \varphi(x)dx = -\int_\R \varphi'(x) \P(X(0) \leq x,L(0)=j)dx,
$$
where $n(x)$ is the normal distribution given by Equation~\eqref{normal}.

Throughout this paper, we shall use the following notation. The functional space
$$
L^2(\R,n) = \left\{ f : \R \to \R : \int_\R f(x)^2 n(x) dx <\infty\right\}
$$
is a Hilbert space equipped with the scalar product defined for $f,g \in L^2(\R,n)$ by
$$
(f,g)_2 = \int_\R f(x)g(x)n(x)dx
$$
and the norm of an element $f \in L^2(\R,n)$ is $\|f\|_2 = \sqrt{(f,g)_2}$. If $\mathcal{H}$ is a separable Hilbert space equipped with the scalar product $(.,.)_\mathcal{H}$ and associated norm $\|.\|_\mathcal{H}$, we define the Hilbert space 
$$
L^2(\R,n;\mathcal{H}) = \left\{ (f_j(x),j \geq 0)\in L^2(\R;n)^\N : \int_\R \|f(x)\|^2_\mathcal{H} n(x)dx <\infty \right\}
$$
equipped with the scalar product
$$
(f,g) = \int_\R (f(x),g(x))_\mathcal{H} n(x)dx .
$$
Finally,  let $\ell^2(\rho)$ be the Hilbert space composed of those sequences $(c_j, j\geq 0)$ taking values in $\R$ and such that $\sum_{j = 0}^\infty c_j^2 \rho^j <\infty$, and equipped with the scalar product defined by: if $c= (c_j)$ and $d = (d_j)$ in $\ell^2(\rho)$, $(c,d)_\rho = \sum_{j=0}^\infty c_j {d}_j \rho^j$;  
the associated norm is defined by: for $c \in \ell^2(\rho)$, $\|c\|_\rho = \sqrt{(c,c)_\rho}$. Let $e_j$ denote the sequence with all entries equal to 0 except the $j$th one equal to 1. The family $(e_j, j \geq 0)$ is a basis for $\ell^2(\rho)$. The space $\ell^2_1(\rho)$ denotes  the subspace of $\ell^2(\rho)$ spanned by the vectors $e_j$ for $j \geq 1$.

In a first step, we determine the infinitesimal generator of the Markov process  $(X(t),L(t))$  taking values in $\R\times \N$. We specifically have the following result.

\begin{lemma}
The process $(X(t),L(t))$ is a Markov process in  $\R\times \N$ with infinitesimal generator $\mathcal{G}$ defined by
\begin{multline}\label{pinot}
\mathcal{G} f(x,j) =\frac{\sigma ^2}{2}\frac{\partial^2 f}{\partial x^2}(x,j)-\alpha (x-m) \frac{\partial  f}{\partial x} (x,j)\\ +\lambda \left(f(x,j+1)-f(x,j)\right)+\mu \phi (x) \ind_{\{j>0\}}\left(f(x,j-1)-f(x,j)\right),
\end{multline}
for every function $f(x,j)$ from $\R\times\N$ in $\R$, twice differentiable with respect to the first variable.
\end{lemma}

\begin{proof}
According to  Equation~\eqref{itosde}, the infinitesimal generator of an
Ornstein-Uh\-len\-beck process applied to some twice differentiable  function $g$ on $\R$ is given by
\begin{equation}
\label{genOU}
Hg = \frac{\sigma ^2}{2}\frac{\partial^2 g}{\partial x^2}(x)-\alpha (x-m) \frac{\partial  g}{\partial x} (x).
\end{equation}
The second part of Equation~\eqref{pinot} corresponds to the infinitesimal generator of
the number of customers in a classical $M/M/1$ queue with arrival rate $\lambda$ and
service rate $\mu \phi (x)$,  when the Ornstein-Uhlenbeck process is in state $x$.  
\end{proof}

For finite sequences $f = (f_j(x))$ of infinitely differentiable functions with compact support, the equation for the invariant measure for the Markov process $(L(t),X(t))$ is given by $\sum_{j\geq 0}\int_{\R}  \mathcal{G} f(x,j)p_j(x)n(x)dx =0$, that is,
\begin{multline*}
\sum_{j\geq 0} \int_{-\infty}^\infty \left(\frac{\sigma^2}{2} \frac{d^2 f_j}{dx^2} -\alpha(x-m)\frac{d f_j}{dx} \right.\\
\left. +\mu\ind_{\{j>0\}} \phi(x) f_{j-1}(x) -(\lambda+\mu\phi(x) \ind_{\{j>0\}})f_j(x) +\lambda f_{j+1}(x)\frac{}{}\right)   {p_j}(x) n(x)dx = 0.
\end{multline*}
\cbstart Via integration by parts, we obtain for every finite sequence $f = (f_j(x))$ of infinitely differentiable functions with compact support
\begin{multline*}
\sum_{j\geq 0} \left(\int_{-\infty}^\infty  \left(\frac{\sigma^2}{2} \frac{d^2 P_j}{dx^2} -\alpha(x-m)\frac{d P_j}{dx} \right) f_j(x) n(x)dx \right.\\
+  \left. \int_{-\infty}^\infty  \left( \mu\ind_{\{j>0\}} \phi(x) f_{j-1}(x)  - (\lambda+\mu\phi(x) \ind_{\{j>0\}})f_j(x) +\lambda f_{j+1}(x)\right)  {p_j}(x) n(x)dx \right)\\ =0
\end{multline*}
and then
\begin{multline*}
\sum_{j\geq 0} \int_{-\infty}^\infty  \left( \frac{\sigma^2}{2} \frac{d^2 {P}_j}{dx^2} -\alpha(x-m)\frac{d {P}_j}{dx} \right. \\ +\left. \mu\ind_{\{j>0\}} P_{j-1}(x) -(\lambda+\mu\phi(x)\ind_{\{j>0\}} )P_j(x) +\lambda \phi(x) P_{j+1}(x) \frac{}{} \right)f_j(x) n(x)dx= 0.
\end{multline*}
This implies the following result.
\cbend

\begin{proposition}
The family $(P_j(x) \stackrel{\mathrm{def}}{=}{p}_j(x)/\rho^j, j \geq 0) \in \mathcal{D}'(\R)^\N$ is solution  \sout{in the sense of distributions} to  the following infinite differential system: for $j \geq 0$, 
\begin{multline}
\label{equilibrsys}
\frac{\sigma^2}{2} \frac{d^2 {P}_j}{dx^2} -\alpha(x-m)\frac{d {P}_j}{dx} \\ +\mu \ind_{\{j>0\}}P_{j-1}(x) -(\lambda+\mu\phi(x)\ind_{\{j>0\}} )P_j(x) +\lambda \phi(x) P_{j+1}(x) = 0.
\end{multline}
\end{proposition}

\subsection{Additional properties}

For the system considered in this paper, we have $\mu\phi(x)>\mu(1- a)$ for all $x \in \R$ and $0<a<1$. Classical stochastic ordering arguments imply  that the process $(L(t))$ is stochastically dominated for the strong ordering sense by the queuing process of the $M/M/1$ queue with input rate $\lambda$ and service rate $\mu(1-a)$. Hence, if the solution $(P_j(x), j \geq 0)$ of the infinite differential system~\eqref{equilibrsys} is related  to the conditional probability density functions $({p}_j(x), j \geq 0)$ of the couple  $(X(0),L(0))$ as $P_j(x) =p_j(x)/\rho^j$ for all $j\geq 0$, then
\begin{equation}
\label{stochordre}
\forall x \in \R, \forall j \geq 0, \quad P_j(x) \leq \frac{1}{(1- a)^j}.
\end{equation}
If $a <1-\sqrt{\rho}$, it is easily checked that for all $x \in \R$, the sequence $(P_j(x), j \geq 0)$ is in the Hilbert space $\ell^2(\rho)$. 

In addition, for all $j \geq 0$
\begin{equation}
\label{ineqtech0}
\int_\R {p}_j(x)^2 n(x) dx \leq \P(L(0)=j)<\infty,
\end{equation}
since ${p}_j(x)= \P(L(0)=j~|~X(0)=x) \leq 1$. It follows that for all $j\geq 0$, the function $p_j(x)$ should be in the space $L^2(\R,n)$. 

Hence, if the solution $(P_j(x), j \geq 0)$ of the infinite differential system~\eqref{equilibrsys} is related to the conditional probability density functions $({p}_j(x), j \geq 0)$ of the couple  $(X(0),L(0))$  as specified above, then $P_j(x) \in L^2(\R,n)$ for all $j \geq 0$. From inequality~\eqref{ineqtech0}, we also deduce that if $a  <1-\sqrt{\rho}$,
$$
\sum_{j=0}^\infty \int_\R \left(\frac{{p}_j(x)}{\rho^j}\right)^2 n(x) dx \rho^j <\infty
$$
since $\P(L(0)=j) \leq \rho^j/(1-a)^j$.

If follows from the above remarks that to show the regularity of the conditional probability density functions $p_j(x)$ for $j \geq 0$ under the assumption $a  <1-\sqrt{\rho}$, we are led to prove  that the differential systems admits a unique regular solution in the space $L^2(\R,n;l^2(\rho))$.

In the next section, we review some properties of the operators associated with the generators of the Ornstein-Uhlenbeck process and the Markov process describing the number of customers in an $M/M/1$ queue.

\section{Some results on the operators associated with the generators of the Ornstein-Uhlenbeck process and the $M/M/1$ queue}
\label{generator}

It is well known in the literature (see for instance \cite{Schaumburg}) that the operator $H$ defined by Equation~\eqref{genOU} is selfadjoint in the Hilbert space $L^2(\R,n)$. The eigenvalues of this  operator are the numbers $-\alpha j$, $j \geq 0$, and the normalized eigenvector  associated with the eigenvalue $-\alpha j$ is  the function $h_j$ given by
\begin{equation}
\label{defhj}
h_j(x) =\frac{1}{\sqrt{2^j j! \sqrt{\pi}}}H_j(\sqrt{2\alpha}(x-m)/\sigma),
\end{equation}
where $H_j(x)$ is the $j$th Hermite polynomial. The sequence $(h_j , j\geq 0)$ is an orthonormal basis of $L^2(\R,n)$. The domain of the operator $H$ is the set
$$
D(H) = \left\{f \in H^2(\R,n) : x^2f \in L^2(\R,n)  \right\},
$$
where $H^2(\R,n)$ is the Sobolev space defined as follows:
$$
H^2(\R,n) = \{f \in C^1(\R) : f, \; f' \in L^2(\R,n) \; \mbox{and the weak derivative } f'' \in L^2(\R,n)\}.
$$

\cbstart \sout{While the operator $H$ is well known in the literature, less information is available on t} The operator $A$ associated with the Markov process describing the number of customers in an $M/M/1$ queue and defined in $\ell^2(\rho)$ by the infinite matrix
\begin{equation}
\label{defmatA}
A=\begin{pmatrix}
-\lambda     &        \lambda          &0                           & .                              & .                &.                                  \\
\mu            & -(\lambda+\mu) & \lambda             & 0                            & .                & .                      \\
0                   &         \mu             & -(\lambda+\mu) & \lambda              & 0               & .            \\
0                   & 0                            &           \mu             & -(\lambda +\mu) & \lambda & . \\
.                   & .                            &           .            & .&. & . 
\end{pmatrix}.
\end{equation}
has already been studied in the technical literature, notably in \cite{Karlin58} (see also \cite{Karlin59}). From these references, we know that the operator $A$ is selfadjoint. Associated with  the  operator $A$ is the operator $A_1$ defined in $\ell^2_1(\rho)$ by the infinite matrix given by
\begin{equation}
\label{defmatA1}
A_1=\begin{pmatrix}
-(\lambda+\mu)     &        \lambda          &0                           & .                              & .                &.                                  \\
\mu            & -(\lambda+\mu) & \lambda             & 0                            & .                & .                      \\
0                   &         \mu             & -(\lambda+\mu) & \lambda              & 0               & .            \\
0                   & 0                            &           \mu             & -(\lambda +\mu) & \lambda & . \\
.                   & .                            &           .            & .&. & . 
\end{pmatrix}.
\end{equation}
Note that the above matrix is the generator of the Markov process describing the number of customers in an $M/M/1$ queue and absorbed at state 0 (see \cite{Karlin58}). Finally, let $A_1^{[N]}$ denote the truncated operator associated with the finite matrix 
\begin{equation}
\label{defmatA1N}
A_1^{[N]}=\begin{pmatrix}
-(\lambda+\mu)     &        \lambda          &0                           & .                              & .                &.                                  \\
\mu            & -(\lambda+\mu) & \lambda              & 0                            & .                & .                      \\
0                   &         \mu             & -(\lambda+\mu) & \lambda              & 0               & .            \\
0                   & 0                            &                      0  & \ddots & \ddots & . \\
.                   & .                            &           .            & .&\mu & -\mu
\end{pmatrix}.
\end{equation}
The above matrix  is the generator of the Markov process describing the number of customers in the finite capacity $M/M/1/N$ queue and absorbed at state 0. (See \cite{Karlin65} for a related model.) In the following, we recall the spectral properties of the operators  $A$, $A_1$ and $A_1^{[N]}$; see \cite{Karlin58,Karlin59,Karlin65} for details
\cbend 

\begin{lemma}
\label{lemtechA}
The operator $A$ in $\ell^2(\rho)$ is bounded and symmetric, and then selfadjoint. In particular, for all $f \in \ell^2(\rho)$, 
\begin{equation}
\label{Amonotone}
0\leq (-Af,f)_\rho\leq \mu (1+\sqrt{\rho})^2\|f\|^2_\rho,
\end{equation}
which implies that the operator $-A$ is monotonic (i.e., $(-Af,f)\geq 0$ for all $f \in \ell^2(\rho)$).
There exists a unique normalized measure $d\psi(z)$, referred to as spectral measure, whose support is the spectrum 
$\sigma(A)$ of operator $A$,  and a family  of spaces  $\{\mathcal{H}_z(\rho)\}$, $z \in \sigma(A)$, such that 
\begin{itemize}
\item the Hilbert space $\ell^2(\rho)$ is equal to the direct sum of the spaces $\mathcal{H}_z(\rho)$, i.e., every $f \in\ell^2(\rho)$ can be decomposed into a family $(f_z, z\in \sigma(A))$, where $f_z \in \mathcal{H}_z(\rho)$ and $\int \|f_z\|^2_\rho d\psi(z) <\infty$. Moreover,
$$
(f,g)_\rho = \int (f_z,g_z)_\rho d\psi(z).
$$
\item The operator $A$ is such $(Af)_z= zf_z$ for $z\in \sigma(A)$, where $(Af)_z$ is the projection of $(Af)$ on the space $\mathcal{H}_z(\rho)$.
\end{itemize}
The spectral measure $d\psi(x)$ is specifically given by
\begin{multline}
\label{mesurespec}
\int h(x)d\psi(x) = (1-\rho)h(0)\\ - \frac{\sqrt{\rho}}{\pi}\int_{\displaystyle-\mu (1+\sqrt{\rho})^2}^{\displaystyle-\mu (1-\sqrt{\rho})^2} \frac{h(x)}{x} \sqrt{1-\left(\frac{x+\lambda+\mu}{2\sqrt{\lambda\mu}}\right)^2}\,dx,
\end{multline}
for any smooth  function $h$. 

The spectrum of the operator $A$ is $\sigma(A) = [-\mu (1+\sqrt{\rho})^2, -\mu (1-\sqrt{\rho})^2]\cup \{0\}$.  The operator $A$ has a unique eigenvalue equal to  $0$, and the eigenspace  $\mathcal{H}_0(\rho)$ is spanned by the vector  $e$ with all components equal to 1. For $z \in (-(\sqrt{\lambda}+\sqrt{\mu})^2,-(\sqrt{\lambda}-\sqrt{\mu})^2)$, the space
$\mathcal{H}_z(\rho)$ is the vector space spanned by the vector $Q(z)$, whose components $Q_j(z)$, $j \geq 0$ are  defined by the following  recursion: 
\begin{equation}
\label{defQj}
\left\{
\begin{array}{l}
{Q}_0(z)=1, \quad {Q}_1(z) = (z+\lambda)/\lambda\\ \\
\mu Q_{j+1}(z) -(z+\lambda+\mu)Q_j(z)+\mu Q_{j-1}(z)=0, \quad j\geq 1.
\end{array}
\right.
\end{equation}
The vectors $(Q(z))$ for $z \in (-(\sqrt{\lambda}+\sqrt{\mu})^2, -(\sqrt{\mu}-\sqrt{\lambda})^2)$ form an orthogonal family with weight function  $d\psi(z)$: for all $j,k$,
$$
\int_\R Q_j(x)Q_k(x)d\psi(x) = \frac{1}{\rho^j}\delta_{j,k},
$$
where $\delta_{j,k}$ is the Kronecker symbol, equal to 1 if $j=k$ and 0 if $j\neq k$.
\end{lemma}

\cbstart Note the polynomials $Q_j(x)$ appearing in the above result are known as perturbed Chebyshev polynomials in the literature on orthogonal polynomials \cite{Sansigre}. \cbend For the operator $A_1$, we have the following result, where we use Chebyshev polynomials of the second kind $(U_n(x))$ defined by the recursion
\begin{equation}
\label{defChebyshev}
\left\{
\begin{array}{l}
{U}_0(x)=1, \quad {Q}_1(x) = 2x \\ \\
 U_{j+1}(x)= 2x U_j(x) - U_{j-1}(x), \quad j\geq 1.
\end{array}
\right.
\end{equation}

\begin{lemma}
\label{lemtechA1}
The operator $A_1$ in the subspace $\mathrm{span}(e_j, j\geq 1)$ is bounded and symmetric, and then selfadjoint. In particular, for all $f \in \mathrm{span}(e_j, j\geq 1)$,
\begin{equation}
\label{coercive}
\mu (1-\sqrt{\rho})^2\|f\|^2_\rho \leq (-A_1f,f)_\rho\leq \mu (1+\sqrt{\rho})^2\|f\|^2_\rho.
\end{equation}
The associated normalized spectral measure $d\psi_1(z)$ is given by
\begin{equation}
\label{mesurespec1}
d\psi_1(x) = \frac{2}{\pi} \sqrt{1-\left(\frac{x+\lambda+\mu}{2\sqrt{\lambda\mu}}\right)^2}\ind_{\{x\in (-\mu(1+\sqrt{\rho})^2,-\mu(1-\sqrt{\rho})^2)\}}\,\frac{dx}{2\sqrt{\lambda\mu}}.
\end{equation}
The spectrum of the operator $A_1$ is diffuse (there are  no eigenvalues) and equal to the interval $\sigma(A_1) = [-\mu(1+\sqrt{\rho})^2, -\mu(1-\sqrt{\rho})^2]$.  The Hilbert space $\mathrm{span}(e_j, j\geq 1)$ is equal to the direct sum of the spaces $\mathcal{H}^{(1)}_z(\rho)$, for $z \in (-\mu (1+\sqrt{\rho})^2,-\mu (1-\sqrt{\rho})^2)$, where the space $\mathcal{H}^{(1)}_z(\rho)$ is the vector space spanned by the vector $Q^{(1)}(z)$, whose components $Q^{(1)}_j(z)$, $j \geq 0$ are  defined by the following  recursion: 
\begin{equation}
\label{defQj1}
\left\{
\begin{array}{l}
{Q}^{(1)}_0(z)=1, \quad {Q}^{(1)}_1(z) = (z+\lambda+\mu)/\lambda\\ \\
\mu Q^{(1)}_{j+1}(z) -(z+\lambda+\mu)Q^{(1)}_j(z)+\mu Q^{(1)}_{j-1}(z)=0, \quad j\geq 1.
\end{array}
\right.
\end{equation}
The vectors $(Q^{(1)}(z))$ for $z \in (-(\sqrt{\lambda}+\sqrt{\mu})^2, -(\sqrt{\mu}-\sqrt{\lambda})^2)$ form an orthogonal family with weight function  $d\psi_1(z)$. The polynomials $(Q_j^{(1)}(z))$ are related to Chebyshev polynomials as follows: for $j\geq 0$,
$$
Q_j^{(1)}(z)= \frac{1}{\rho^{j/2}} U_j\left(\frac{z+\lambda + \mu}{2\sqrt{\lambda\mu}}  \right).
$$
\end{lemma}

Finally, for the operator $A_1^{[N]}$, we have the following result.

\begin{lemma}
\label{reftechA1N}
The operator $A_1^{[N]}$ is symmetric (and then selfadjoint) in the vector space $\mathrm{span}(e_1, \ldots,e_N)$ equipped with the scalar product induced by $(.,.)_\rho$. The eigenvalues of the operator  $A_1^{[N]}$ are the solutions to  the polynomial equation
$$
Q^{(1)}_{N+1}(-x;1)= Q^{(1)}_N(-x;1),
$$
where the polynomials $Q_j^{(1)}(x)$ are defined by the recursion~\eqref{defQj1}. The eigenvalues are denoted by $-x^{[N]}_j$, $j=1, \ldots, N$ with $ x_1^{[N]} < x_2^{[N]}< \ldots <x_N^{[N]}$. The vectors $Q^{(1,N)}(x_j)$, $j =1,\ldots,N$, form an orthogonal basis of $\mathrm{span}(e_1,\ldots,e_N)$, where $Q^{(1,N)}(x_j)$ is the vector with the $k$th component equal to $Q^{(1)}_k(x_j)$, $k,j = 1,\ldots,N$, where the polynomials $(Q_k^{(1)}(z))$ are defined by Equation~\eqref{defQj1}.
\end{lemma}

The operator $A$ naturally induces in $L^2(\R,n;l^2(\rho))$ an operator that we still denote by $A$. The same property is valid for the operators $A_1$ and $A_1^{[N]}$ in the spaces $L^2(\R,n;\mathrm{span}(e_j,j \geq 1))$ and $L^2(\R,n;\mathrm{span}(e_j,j = 1,\ldots,N))$, respectively. Similarly, the operator $H$ induces in $L^2(\R,n;l^2(\rho))$ an operator that we  still denote by $H$ and which is defined as follows: for $f \in L^2(\R,n;l^2(\rho))$, $Hf$ is the element with the $j$th component equal to  $Hf_j$. This operator also induces in $L^2(\R,n;\mathrm{span}(e_j,j \geq 1))$ and $L^2(\R,n;\mathrm{span}(e_j,j = 1,\ldots,N))$  operators denoted by $H_1$ and $H_1^{[N]}$, respectively. The operators $A$, $A_1$, $A_1^{[N]}$, $H$, $H_1$ and $H_1^{[N]}$ are clearly selfadjoint in the spaces where they are defined.

With the above definitions, the fundamental differential system~\eqref{equilibrsys}  reads
\begin{equation}
\label{fokker}
(H+A)f +Vf=0,
\end{equation}
where the operator $V$ is defined by: for $f \in L^2(\R,n;l^2(\rho))$, $Vf = (\phi(x)-1)Bf$, where $B$ is the operator associated with the infinite matrix
$$
B=\begin{pmatrix}
0     &        \lambda        &0                           & .                              & .   & .     & .                                  \\
0              & -\mu & \lambda            & 0                            & .                & .      & .                        \\
0                   &         0              & -\mu  & \lambda          & 0               & .       & .   \\
0                   & 0                            &           0            & -\mu & \lambda &0 & .\\
.                   & .                            &           .            & .&. & . & .
\end{pmatrix}.
$$
\cbstart    In the notation of Equation~\eqref{eqbase}, we have $\Omega = H+A$ and $V(\varepsilon)= -\varepsilon ((x\wedge (a/\varepsilon))\vee (-b/\varepsilon))B$.\cbend  

The matrices 
$$
B_1=\begin{pmatrix}
-\mu     &        \lambda        &0                           & .                              & .   & .     & .                                  \\
0              & -\mu & \lambda            & 0                            & .                & .      & .                        \\
0                   &         0              & -\mu  & \lambda          & 0               & .       & .   \\
0                   & 0                            &           0            & -\mu & \lambda &0 & .\\
.                   & .                            &           .            & .&. & . & .
\end{pmatrix}.
$$
and 
$$
B_1^{[N]}=\begin{pmatrix}
-\mu     &        \lambda          &0                           & .                              & .                &.                                  \\
0           & -\mu & \lambda              & 0                            & .                & .                      \\
0                   & 0                            &                      \ddots  & \ddots & \ddots & . \\
.                   & .                            &           .            & .& 0 & -\mu
\end{pmatrix}
$$
define the operators $B_1$ and $B_1^{[N]}$ in the spaces $L^2(\R,n;\mathrm{span}(e_j,j \geq 1))$ and $L^2(\R,n;\mathrm{span}(e_j,j = 1,\ldots,N))$, respectively. It is straightforwardly checked that the operator $B$, $B_1$ and $B^{[N]}_1$ are bounded with a norm less than or equal to  $\mu (1+\sqrt{\rho})$. With the operators $B_1$ and $B_1^{[N]}$ are associated the operators $V_1$ and $V_1^{[N]}$ induced by $V$ in the spaces $L^2(\R,n;\mathrm{span}(e_j,j \geq 1))$ and $L^2(\R,n;\mathrm{span}(e_j,j = 1,\ldots,N))$, respectively.

In the next section, we prove that the differential system~\eqref{equilibrsys} (or equivalently Equation~\eqref{fokker}) has a unique solution in $L^2(\R,n;\ell^2(\rho))$.

\section{Existence and uniqueness of a solution}
\label{existence}

When we  refer to the existence of  a density probability density  function satisfying the
differential system~\eqref{equilibrsys}, we  think of a vector $(P_j(x))$  such that every
function  $P_j(x)$ is  twice  continuously  differentiable over  $\R$  (i.e., $P_j(x)  \in
C^2(\R)$ for all $j \geq 0$). But,  this differential system may have a solution, which is
in $L^2(\R,n;\ell^2(\rho))$  but with  components not in  $C^2(\R)$. In the  following, we
prove  that   the  differential  system~\eqref{equilibrsys}  has  a   unique  solution  in
$L^2(\R,n;\ell^2(\rho))$ and then we show at the end of the section that the components of
the solution are $C^2(\R)$ functions.

If $f \in L^2(\R,n;\ell^2(\rho))$ is solution to  Equation~\eqref{fokker}, then 
\begin{equation}
\label{fokkerops1}  
(H_1 + A_1) f^1 + V_1 f^1 =  -\mu e_1(f_0),
\end{equation}
where $f^1$ is the projection of $f$ on the space $L^2(\R,n,\mathrm{span}(e_j,j\geq 1))$ and $e_1(f_0)$ is the element of $L^2(\R,n,\mathrm{span}(e_j,j\geq 1))$  with the first component equal to $f_0(x)$ and all other components equal to 0.

\cbstart The operator $(H_1+A_1)$ is self-adjoint and invertible. It is straightforwardly checked that the norm 
\begin{multline*}
\|(H_1+A_1)^{-1}\| \stackrel{def}{=}\\  \sup\{\left\|(H_1+A_1)^{-1}f\right\| : f \in L^2(\R,n,\mathrm{span}(e_j,j\geq 1)), \|f\|=1\}  
\end{multline*}
is such that 
$$
\|(H_1+A_1)^{-1}\| \leq \frac{1}{\mu(1-\sqrt{\rho})^2},
$$
since by Lemma~\ref{lemtechA1} and the monotonicity of the operator $-H_1$, we have for all $f$ in $L^2(\R,n,\mathrm{span}(e_j,j\geq 1))$
$$
(-(H_1+A_1)f,f) \geq (-H_1f,f) +(-A_1f,f) \geq -\mu(1-\sqrt{\rho})^2\|f\|^2.
$$
\cbend

In addition, if $f \in L^2(\R,n;\ell^2(\rho))$ is a solution to Equation~\eqref{fokker}, then the function $\sum_{j=0}^\infty \rho^j f_j(x) \in L^2(n)$ since by Schwarz inequality
$$
\int_{-\infty}^\infty \left| \sum_{j=0}^\infty \rho^j f_j(x)\right|^2n(x)dx \leq \frac{1}{1-\rho} \int_{-\infty}^\infty \sum_{j=0}^\infty |f_j(x)|^2 \rho^j n(x)dx = \frac{1}{1-\rho} \|f\|^2<\infty.
$$
By summing all the lines of Equation~\eqref{fokker}, we see that the function $\sum_{j=0}^\infty \rho^j f_j(x) $ has to be solution to the equation $H g=0$ in $L^2(\R,n)$.

By using the above observations, we prove the existence and uniqueness of a non trivial solution in $L^2(\R,n;\ell^2(\rho))$ to Equation~\eqref{fokker} by showing the following results:
\begin{enumerate}
\item If $\varepsilon$ satisfies some condition, Equation~\eqref{fokkerops1} has a solution for all $f_0 \in L^2(n)$; if $(f_j)$ is the solution, we set
\begin{equation}
\label{defK}
Kf_0 = \sum_{j=1}^\infty \rho^j f_j \in L^2(\R,n).
\end{equation}
\item If $f$ is a non trivial solution to Equation~\eqref{fokker}, then $f_0$ is a non trivial solution to equation
\begin{equation}
\label{eqf0}
f_0 + Kf_0 = 1.
\end{equation}
\item Equation~\eqref{eqf0} has a unique non trivial solution.
\end{enumerate}
To prove the last point, we intend to use the Fredholm alternative, which requires that the operator $K$ is compact.

\begin{lemma}
Under the condition
\begin{equation}
\label{condeps}
a\vee b < \frac{(1-\sqrt{\rho})^2}{1+\sqrt{\rho}},
\end{equation}
where $a\vee b=\max(a,b)$, Equation~\eqref{fokkerops1} has a unique solution in $L^2(\R,n;\ell_1^2(\rho))$.
\end{lemma}

\begin{proof}
Equation~\eqref{fokkerops1} can be rewritten as
$$
(\I + (H_1+A_1)^{-1}V_1)f = -\mu(H_1+A_1)^{-1}e_1(f_0).
$$
Since $V_1$ is bounded so that for all $f \in L^2(\R,n;\ell_1^2(\rho))$, $|(V_1f,f)| \leq \mu  (a\vee b) (1+\sqrt{\rho}) \|f\|^2$, we deduce that the operator $(H_1+A_1)^{-1}V_1$ is bounded with a norm less than or equal to  $(a\vee b) (1+\sqrt{\rho})/ (1-\sqrt{\rho})^2$. Under Condition~\eqref{condeps}, the norm of operator $(H_1+A_1)^{-1}V_1$ is less than 1 and we then deduce that $(\I + (H_1+A_1)^{-1}V_1)$ is invertible \cite{Reed} and Equation~\eqref{fokkerops1} has a unique solution in $L^2(\R,n;\ell_1^2(\rho))$.
\end{proof}

It is worth noting that under Condition~\eqref{condeps}, we have $a  <(1-\sqrt{\rho})$. Moreover, the above result ensures that the operator $K$ is well defined by Equation~\eqref{defK}. Now, we prove that under the same condition, the operator $K$ is monotonic.

\begin{lemma}
Under Condition~\eqref{condeps}, the operator $K$ is monotonic, which implies that there exists at most one  non trivial solution to Equation~\eqref{eqf0}.  
\end{lemma}

\begin{proof}
We first note that from Equation~\eqref{defK}, we have
$$
(Kf_0,f_0)_2 = (-\mu (H_1+A_1 +V_1)^{-1}e_1(f_0),e(f_0)),
$$
where $e(f_0)$ is the vector with all entries equal to $f_0$. The above equation can be rewritten as
$$
(Kf_0,f_0)_2 = (-\mu (\I +A_1^{-1}(H_1+V_1))^{-1}A_1^{-1}e_1(f_0),e(f_0)).
$$
Since $-\mu A_1^{-1}e_1(f_0) = e(f_0)$, we have
\begin{equation}
\label{ineqtemp}
(Kf_0,f_0)_2 = ((\I +A_1^{-1}(H_1+V_1))^{-1}e(f_0),e(f_0)) 
\end{equation}
and hence $(Kf_0,f_0)_2\geq 0$. Indeed, for all $f \in L^2(\R,n;\ell^2_1(\rho))$
\begin{eqnarray*}
((\I + A_1^{-1}H_1 +A_1^{-1}V_1)f,f)   &=& ((\I + A_1^{-1}H_1)f,f) +(A_1^{-1}V_1f,f) \\
&\geq& \left(1- \frac{(a\vee b)(1+\sqrt{\rho})}{(1-\sqrt{\rho})^2}\right) \|f\|^2,
\end{eqnarray*}
where we have used the fact that the operator $A_1^{-1}H_1$ is monotonic, $\|V_1\|\leq \mu(a\vee b)(1+\sqrt{\rho})$, and $\|A_1^{-1}\|\leq 1/(\mu(1-\sqrt{\rho})^2)$. The above inequality  implies that 
\begin{multline*}
((\I + A_1^{-1}H_1 +A_1^{-1}V_1)^{-1}f,f) \\ \geq \left(1- \frac{( a\vee b) (1+\sqrt{\rho})}{(1-\sqrt{\rho})^2}\right) \|(\I + A_1^{-1}H_1 +A_1^{-1}V_1)^{-1}f\|^2 \geq 0,
\end{multline*}
and Inequality~\eqref{ineqtemp} follows.
\end{proof}

We now turn to the compactness of the operator $K$. The major difficulty comes from the fact that the operator $(H_1+A_1)^{-1}$ is not compact, since we know that the spectrum of this self-adjoint operator is not discrete.  However, by truncating the infinite matrix defined by Equation~\eqref{defmatA}, we can introduce compact operators and subsequently prove  that the operator $K$ is compact. Let us fix some $N>0$. We first prove the following technical lemma.

\begin{lemma}
For $N>0$, the operator $(H^{[N]}_1+A^{[N]}_1)^{-1}$ in $L^2(\R,n; \mathrm{span}(e_1,\ldots,e_N))$ is compact.
\end{lemma}

\begin{proof}
By using Lemma~\ref{reftechA1N} and the orthonormal basis $(h_n)$ is $L^2(\R,n)$, we know that the family $e_{j,k}(x) = h_k(x)Q^{(1,N)}(x_j)$ for $k \geq 0$ and $j=1,\ldots, N$ forms an orthogonal basis of $L^2(\R,n;\mathrm{span}(e_1,\ldots,e_N))$. In particular, we have
$$
(H^{[N]}_1+A{[N]}_1)^{-1}e_{j,k}(x) = -\frac{1}{k\alpha + x^{[N]}_j} e_{j,k}(x).
$$
The operator $(H^{[N]}_1+A^{[N]}_1)$ then appears as the norm limit as $M\to \infty$ of the finite  rank operators $(H^{[N,M]}_1 + A^{[N,M]}_1)^{-1}$  defined in the vector space $\mathrm{span}(e_{j,k}(x),j=1,\ldots,N,k=0,\dots,M)$ by
$$
(H^{[N,M]}_1 + A^{[N,M]}_1)^{-1} e_{j,k}(x) = -\frac{1}{\alpha k + x^{[N]}_j}  e_{j,k}(x)
$$
and the result follows. 
\end{proof}

\begin{lemma}
Under Condition~\eqref{condeps}, the operator $K$ is compact.
\end{lemma}

\begin{proof}
Let us consider a bounded sequence $(f^i_0)$ in $L^2(\R,n)$. (Without loss of generality, we assume that $\|f^i_0\|_2 = 1$.)  Since the operator  $(H^{[N]}_1 + A^{[N]}_1)^{-1} V^{[N]}_1$ is bounded with a norm less than or equal to $(a\vee b)\mu (1+\sqrt{\rho})/x_1^{[N]}$, and the operator $(H^{[N]}_1+A^{[N]}_1)^{-1}$ is compact, we deduce that the operator $(H_1^{[N]} + A_1^{[N]}+ V_1^{[N]})^{-1}$ is compact. 

Let $f^i = (f^i_j(x))$ denote the vector $-\mu(H_1+A_1+V_1)^{-1}e_1(f^{i}_0)$. Since the
operator $(H_1+A_1+V_1)^{-1}$ is bounded with a norm less than or equal to
$1/(\mu((1-\sqrt{\rho})^2- a\vee b \varepsilon (1+\sqrt{\rho})))$, the vector $(f^i)$ is
such that 
\begin{equation}
\label{bornetech}
\sqrt{\sum_{j=1}^N \|f^{i}_j\|^2_2 \rho^j} \leq \|f^i\|  \leq \frac{1}{((1-\sqrt{\rho})^2- a\vee b  (1+\sqrt{\rho}))}.
\end{equation}

We have
$$
\left(\begin{array}{c} f^{i}_1 \\ \vdots \\f^{i}_N \end{array}\right) = (H^{[N]}_1+A^{[N]}_1+V_1^{[N]})^{-1}\left(\begin{array}{c} -\mu f^{i}_0 \\ 0 \\ \vdots \\ 0  \\ -\lambda \phi f^{i}_{N+1} +\lambda f^{i}_N \end{array}\right)
$$
In view of inequality~\eqref{bornetech}, the sequence appearing in the right hand side of the above equation is bounded. Hence, since the operator $(H^{[N]}_1+A^{[N]}_1+V_1^{[N]})^{-1}$ is compact, it is possible to extract a sub-sequence $(f^{i_k})$ such that 
$$
(H^{[N]}_1+A^{[N]}_1+V_1^{[N]})^{-1}\left(\begin{array}{c} -\mu f^{i_{k}}_0 \\ 0 \\ \vdots \\ 0  \\ -\lambda \phi f^{i_{k}}_{N+1} +\lambda f^{i_{k}}_N \end{array}\right) \to \left(\begin{array}{c}  f^{\infty}_1 \\ f_2^\infty \\ \vdots \\ f_{N-1}^\infty  \\ f^{\infty}_N \end{array}\right)
$$
as $k \to\infty$ in $L^2(\R,n;\mathrm{span}(e_1,\ldots,e_N))$, where the vector appearing in the right hand side of the above equation is in  $L^2(\R,n;\mathrm{span}(e_1,\ldots,e_N))$.

In particular, we have $f^{i_k}_N \to f^{\infty}_N$ in $L^2(\R,n)$ as $k \to \infty$. This implies that 
\begin{multline*}
\left(\begin{array}{c} f^{i_k}_{N+1} \\  f^{i_k}_{N+2} \\   \vdots \end{array}\right) = (H^{[N]}_1+A^{[N]}_1+V_1^{[N]})^{-1}\left(\begin{array}{c} -\mu f^{i_k}_{N} \\ 0 \\ \vdots  \end{array}\right) \\ \to (H^{[N]}_1+A^{[N]}_1+V_1^{[N]})^{-1}\left(\begin{array}{c} -\mu f^{\infty}_N \\ 0 \\ \vdots  \end{array}\right)
\end{multline*}
as $k \to \infty$ in $L^2(\R,n,\ell^2_1(\rho))$, since the operator $(H^{[N]}_1+A^{[N]}_1+V_1^{[N]})^{-1}$ is bounded. We set
$$
\left(\begin{array}{c} f^{\infty}_{N+1} \\ f^\infty_{N+2} \\ \vdots  \end{array}\right) = (H^{[N]}_1+A^{[N]}_1+V_1^{[N]})^{-1}\left(\begin{array}{c} -\mu f^{\infty}_N \\ 0 \\ \vdots  \end{array}\right)
$$
The vector with the $j$th component equal to $f^\infty_j$ is in $L^2(\R,n,\ell^2(\rho))$ and we have $Kf^{i_k}_0\to\sum_{j=1}^\infty \rho^j f^\infty_j$ in $L^2(\R,n)$ as $k \to \infty$. We then deduce that from every bounded sequence $(f^i_0)$ in  $L^2(\R,n)$, we can extract a sub-sequence $(f^{i_k}_0)$ such that $Kf_0^{i_k}$ is converging in $L^2(\R,n)$. The operator $K$ is hence compact.
\end{proof}

By using the above lemmas and the Fredholm alternative, we can state  the following result.

\begin{proposition}
\label{Propalphi}
Under Condition~\eqref{condeps}, Equation~\eqref{eqf0} has a unique solution. This establishes that Equation~\eqref{fokker} has a unique non trivial solution in $L^2(\R,n,\ell^2(\rho))$ if Condition~\eqref{condeps} is satisfied.
\end{proposition}

The above result has been established for the perturbation function $\phi(x) = 1-\varepsilon ((x\wedge (a/\varepsilon))\vee (-(b/\varepsilon)))$ and we have exploited the fact that the function $|1-\phi(x)|$ is bounded by $a\vee b$. In fact, it is possible to prove a similar result when $\phi$ is replaced with $\Phi(x) = 1-\varepsilon x$. Let us define the operator $W$ in $L^2(\R,n;\ell^2(\rho))$ by: if $f= (f_j(x), j\geq 0)$
$$
Wf = -\varepsilon x Bf.
$$
Note that the domain of $W$ is given by
$$
D(W) = \left\{ (f_j(x),j \geq 0) \in L^2(\R,n)^\N : \int_\R\sum_{j=0}^\infty x^2 f_j(x)^2 n(x)dx <\infty\right\}.
$$
\cbstart In the notation of Equation~\eqref{eqbasebis}, we have $V(\Phi) = -xB$.\cbend

We can then state the following result, whose proof is given in Appendix~\ref{demPhi}.

\begin{proposition}
\label{PropalPhi}
Under the condition
\begin{equation}
\label{condeps2}
2  \varepsilon \frac{(1+\sqrt{\rho})}{(1-\sqrt{\rho})^2} \left(m+\frac{\sigma}{\sqrt{\alpha}}\right) <1,
\end{equation}
the equation 
\begin{equation}
\label{fokkerbis}
(H+A+W)f = 0
\end{equation}
has a unique solution in $L^2(\R,n,\ell^2(\rho))$.
\end{proposition}

To prove the existence and the uniqueness of the solutions to Equations~\eqref{fokker} and \eqref{fokkerbis}, we have only supposed that the components of the solutions are in $D(H)$, in particular the  components are  in $H^2(\R,n)$. But, by examining the differential systems satisfied by the different components, notably by taking into account the continuity of the functions $\phi(x)$ and $\Phi(x)$, these components are clearly in $C^2(\R)$.

To conclude this section, let us mention that the solution to the differential system~\eqref{equilibrsys} have the following probabilistic interpretation: for all $j\geq 0$, in the stationary regime $P_j(x) = \P(L(t)=j~|~X(t)=x)/\rho^j$. In particular, we have for all $j\geq 0$
\begin{equation}
\label{positive}
0\leq P_j(x) \leq 1
\end{equation} 
and for all $x\in\R$
\begin{equation}
\label{normalisation}
\sum_{j=0}^\infty P_j(x)\rho^j =1.
\end{equation}

\section{Perturbation analysis}\label{perturbation}

\cbstart
The goal of this section is to prove the following Reduced Service Rate approximation.

\begin{theorem}\label{theo1}
For  sufficiently small $\varepsilon$, the first order expansion of the generating function 
of the stationary distribution of  $(L(t))$ is given by
$$
\mean\left(u^{L(t)}\right) = \frac{1-\rho}{1-\rho u} - \frac{\rho(1-u)}{(1-\rho u)^2} m \varepsilon + o(\varepsilon).
$$
\end{theorem}
Therefore,  $\mean(u^{L(t)}) \sim \mean(u^{L_\varepsilon})$, where
$L_\varepsilon$ has the stationary distribution of the number of customers in an $M/M/1$
queue when the server rate is $1-\varepsilon m$. This shows a principle of reduced service
rate approximation, i.e., everything happens as if the server rate were fixed equal to $1-m\varepsilon$.

To show the above result, we proceed as follows:
\begin{enumerate}
\item We compare the solutions to Equations~\eqref{fokker} and \eqref{fokkerbis} when Conditions~\eqref{condeps} and \eqref{condeps2} are satisfied. In particular, we compute  an upper bound for the norm of their difference.
\item We develop the solution $g$ to Equation~\eqref{fokkerbis} in power series expansion of $\varepsilon$. In particular, we explicitly compute the two first terms.
\item We finally prove Theorem~\ref{theo1}.
\end{enumerate}

Throughout this section, we denote by $P$ and $g$ the solutions to  Equations~\eqref{fokker} and \eqref{fokkerbis} in $L^2(\R,n;\ell^2(\rho))$, respectively. 
\cbend

\subsection{Comparison of the solutions $P$ and $g$}

The solutions to Equations~\eqref{fokker} and \eqref{fokkerbis} when they exist are close
to each other when $\varepsilon$ is small. We specifically have the following result. 

\begin{proposition}
\label{dominationp}
Assume that Conditions~\eqref{condeps} and \eqref{condeps2} are satisfied. These solutions $P$ and $g$ are such that
\begin{equation}
\|P - g\| \leq D(\varepsilon),
\end{equation}
\cbstart where the function $D(\varepsilon)$ is given by
\begin{equation}
\label{defDeps}
D(\varepsilon) = M\left( 1+ \sqrt{\frac{\rho}{1-\rho}}+ \lambda \sqrt{\frac{\rho}{1-\rho}} M\right)\Delta(\varepsilon)
\end{equation}
with 
\begin{eqnarray*}
M & = & \frac{1}{\mu(1-\sqrt{\rho})^2}\left(1+ \frac{(1+\sqrt{\rho})}{(1-\sqrt{\rho})^2} \left(m+\frac{\sigma}{\sqrt{\alpha}}\right) \right), \\
\Delta(\varepsilon)^2 &=& (\mu^2+3\lambda\mu) \int_\R \left((\varepsilon x-a)^2\ind_{\{x\geq a/\varepsilon\}}+ (\varepsilon x+b)^2\ind_{\{x\leq -b/\varepsilon\}}\right) n(x)dx.
\end{eqnarray*}
\cbend
The function $D(\varepsilon)$ is $O\left(\varepsilon^{5/2} \exp(-\alpha(a\wedge b)^2/(2\sigma^2\varepsilon^2))\right)$ when $ \varepsilon \to 0$.
\end{proposition}

\begin{proof}
The vector $P$ is such that
$$
(H+W+A)P+(V-W)P=0
$$
and hence, if $P^1$ denotes the projection of $P$ on the space $\mathrm{span}(e_j, j\geq 1)$,
$$
(H_1+A_1+W_1)P^1 = (W_1-V_1)P^1-\mu e_1(P_0)
$$
and then
\begin{equation}
\label{eqforP1}
P^1 = (H_1+A_1+W_1)^{-1}(W_1-V_1)P^1-\mu (H_1+A_1+W_1)^{-1} e_1(P_0).
\end{equation}
It follows that 
\begin{multline*}
\sum_{j=1}^\infty P_j(x)\rho^j \equiv (P^1,e^1)_\rho = ( (H_1+A_1+W_1)^{-1}(W_1-V_1)P^1,e^1)_\rho  \\ -\mu ((H_1+A_1+W_1)^{-1} e_1(P_0),e^1)_\rho.
\end{multline*}
We have
$$
 -\mu ((H_1+A_1+W_1)^{-1} e_1(P_0),e^1)_\rho = K'g_0,
$$
where the operator $K'$ is defined as the operator $K$ (defined by Equation~\eqref{defK}) but by replacing $V$ with $W$. In addition, since $P$ satisfies $(P,e)_\rho=1$, we come up with the conclusion that $P_0$ verifies
$$
1-P_0 = ( (H_1+A_1+V_1)^{-1}(W_1-V_1)P^1,e^1)_\rho  +K'P_0.
$$
Since $g_0 +K'g_0 = 1$, we obtain
$$
(P_0-g_0) + K'(P_0-g_0) =   ( (H_1+A_1+W_1)^{-1}(W_1-V_1)P^1,e^1)_\rho.
$$
Since the operator $K'$ is monotonic, the above equation implies that 
$$
\|P_0-g_0\|_2 \leq \sqrt{\frac{\rho}{1-\rho}}  \| (H_1+A_1+W_1)^{-1}\| \|(W_1-V_1)P^1\|.
$$

The norm  $\|(W_1-V_1)P^1\|$ is given by
\begin{multline*}
\|(W_1-V_1)P^1\|^2  = \\ \sum_{j=1}^\infty \int_\R \left((\varepsilon x-a)^2\ind_{\{x\geq a/\varepsilon\}}+ (\varepsilon x+b)^2\ind_{\{x\leq -b/\varepsilon\}}\right) (\mu P_j(x) - \lambda P_{j+1}(x))^2 \rho^j n(x)dx
\end{multline*}
and by using Equation~\eqref{positive}, we obtain
\begin{multline*}
\|(W_1-V_1)P^1\|^2  \\ \leq (\mu^2+3\lambda\mu) \int_\R \left((\varepsilon x-a)^2\ind_{\{x\geq a/\varepsilon\}}+ (\varepsilon x+b)^2\ind_{\{x\leq -b/\varepsilon\}}\right) n(x)dx
\end{multline*}
Simple computations show that
\begin{eqnarray*}
 \int_\R (\varepsilon x-a)^2\ind_{\{x\geq a/\varepsilon\}}n(x)dx &\sim & \frac{5\varepsilon^5\sigma^5}{8\alpha^{5/2}a^3\sqrt{\pi}}e^{-\frac{\alpha a^2}{\varepsilon^2\sigma^2}},\\
 \int_\R (\varepsilon x+b)^2\ind_{\{x\leq -b/\varepsilon\}} n(x)dx&\sim & \frac{5\varepsilon^5\sigma^5}{8\alpha^{5/2}b^3\sqrt{\pi}}e^{-\frac{\alpha b^2}{\varepsilon^2\sigma^2}},
\end{eqnarray*}
when $\varepsilon \to 0$. The term $\|(W_1-V_1)P^1\|$   is hence $O(\exp(\varepsilon^{5/2} \exp(-\alpha(a\wedge b)^2/(2\sigma^2\varepsilon^2)))$ when $\varepsilon \to 0$. In addition, since by using Equation~\eqref{normW1} in Appendix~\ref{demPhi}
$$
 \| (H_1+A_1+W_1)^{-1}\| \leq \frac{1}{\mu(1-\sqrt{\rho})^2}\left(1+ \frac{(1+\sqrt{\rho})}{(1-\sqrt{\rho})^2} \left(m+\frac{\sigma}{\sqrt{\alpha}}\right) \right),
$$
we deduce that $\|f_0-g_0\|_2$ is dominated by a term, which is $O(\exp(\varepsilon^{5/2} \exp(-\alpha(a\wedge b)^2/(2\sigma^2\varepsilon^2)))$ when $\varepsilon \to 0$. Finally, by using the fact that $(H^1+A_1+W_1)g= -\mu e_1(g_0)$, we deduce from Equation~\eqref{eqforP1} that
$$
P^1-g^1 =  (H_1+A_1+W_1)^{-1}(W_1-V_1)P^1-\mu (H_1+A_1+W_1)^{-1} e_1(P_0-g_0),
$$
which implies that 
$$
\|P^1-g^1\| \leq  \| (H_1+A_1+W_1)^{-1}\| \|(W_1-V_1)P^1\| + \lambda  \| (H_1+A_1+W_1)^{-1}\| \|P_0-g_0\|_2
$$
and hence $\|P^1-g^1\|$ is dominated by a term which is $O(\exp(\varepsilon^{5/2} \exp(-\alpha(a\wedge b)^2/(2\sigma^2\varepsilon^2)))$ when $\varepsilon \to 0$.
\end{proof}

\subsection{Power series expansion of the solution $g$}

\subsubsection{Notation}

We assume that the solution $g$ to Equation~\eqref{fokkerbis} can be uniquely decomposed as a power series expansion of  the form 
\begin{equation}
\label{expandg}
g = g^{(0)} + \varepsilon g^{(1)} + \varepsilon^2 g^{(2)} + ....,
\end{equation}
where $g^{(i)} \in L^2(\R,n;\ell^2(\rho))$ for $i \geq 0$. In addition, to facilitate the computations, we shall consider  the generating function $g_{u}(x) = \sum_{j=0}^\infty g_j(x)u^j\rho^j$, which is an element of $L^2(\R,n;\ell^2(1/\rho))$. Indeed, if $g$ is written in the form
$$
g = \sum_{j=0}^\infty c_j(u)h_j(x),
$$
where $c_j(u) = \sum_{m=0}^\infty c_{j,m}u^m$ with $(c_{j,m},m\geq 0) \in \ell^2(\rho)$, then $g_u(x)$ can be written as
$$
g_u(x) = \sum_{j=0}^\infty C_j(u)h_j(x),
$$
with $C_j(u) = \sum_{m=0}^\infty C_{j,m}u^m = c_j(\rho u)$. Since $(c_{j,m},m\geq 0)  \in \ell^2(\rho)$, $(C_{j,m},m\geq 0) \in \ell^2(1/\rho)$. Finally, we have $\|g\|^2 = \sum_{j=0}^\infty \|c_j\|^2_\rho =  \sum_{j=0}^\infty \|C_j\|^2_{1/\rho}$.

The generating function $g_u(x)$ will be expanded as
\begin{equation}
\label{expandgu}
g_{u}(x) = g_u^{(0)}(x) + \varepsilon g_u^{(1)}(x) +\varepsilon^2g_u^{(2)}(x) + \cdots, 
\end{equation}
where $g_u^{(i)} \in L^2(\R,n)$ for all $i \geq 0$. The function $g_u^{(0)}(x)$ corresponds to the case $\varepsilon = 0$ and  
is given by
\begin{equation}
\label{independance}
g_u^{(0)}(x)= \frac{c(g)}{1-\rho u},
\end{equation} 
where $c(g)$ is the normalizing constant.

In the following, we  prove that the elements $g^{(i)}$ have to satisfy a
recurrence relation of the form $g^{(i)} = \Theta(x g^{(i-1)})$ for $i  \geq 1$ and for
some linear operator $\Theta$ whose norm is finite. 

In the following, we assume that the expansion~\eqref{expandgu} is valid and we investigate the conditions which have to be satisfied by the elements $g^{(i)}$. In a first step, we prove the following  property satisfied by the functions $(g_u^{(i)}(x))$.
\begin{lemma}
\label{domg}
For $i\geq 0$, the vector $g^{(i)}$ is in $L^2_i(\R,n;\ell^2(\rho))$, where $L^2_i(\R,n;\ell^2(\rho))$ is the sub-space of $L^2(\R,n;\ell^2(\rho)$ composed of those elements $(f_j(x))$ such that $f_j(x) \in \mathrm{span}(h_0,\ldots,h_i)$ for all $j \geq 0$, the functions $h_j$ being defined by Equation~\eqref{defhj}; the  function $g_u^{(i)}(x)$ in Expansion~\eqref{expandgu} hence satisfies  for $N > i$
\begin{equation}
\lim_{x \to \pm \infty} \frac{1}{x^N} g_u^{(i)}(x)  =0.
\end{equation}
\end{lemma}

\begin{proof}
The proof is by  mathematical induction. The result is true for $i=0$ since $g^{(0)}=c(g) e$ ($e$ being the vector with all components  equal to 1).   

If the result is true for $i$. From Equation~\eqref{fokkerbis},
we have 
$$
(H+A)g^{(i+1)} = -Wg^{(i)}.
$$
\cbstart By using the recurrence relation satisfied by Hermite polynomials \cite{Lebedev}
\begin{equation}
\label{recHermite}
H_{j+1}(x)-2xH_j(x)+2iH_{j-1}(x) = 0,
\end{equation}
it is easily checked that the image by the multiplication by $x$ of $\mathrm{span}(h_0,\ldots, h_i)$ is $\mathrm{span}(h_0,\ldots, h_{i+1})$. \cbend Therefore, since by
assumption $g^{(i)}$ belongs to $L^2_i(\R, n;\ell^2(\rho))$, we
immediately deduce from the uniqueness of the decomposition on the basis $(h_ie_j, i \geq 0, j\geq 0)$ of the Hilbert space $L^2(\R,n;\ell^2(\rho))$ and the selfadjointness of the operator $H+A$, 
that $g^{(i+1)}$ is in $L^2_{i+1}(\R,n;\ell^2(\rho))$ and the result follows. 
\end{proof}

\subsubsection{First order term}

In a first step, we pay special attention to the derivation of the first order term
because it gives the basic arguments to derive higher order terms. Moreover, the explicit
form of the first order term will be used to examine the validity of the reduced service
rate approximation (see Theorem~\ref{theo1}). 

On the basis of the domination property given by Lemma~\ref{domg}, we explicitly compute
the function $g^{(1)}_u(x)$. From Equation~\eqref{fokkerbis}, it is easily checked that the
function $g^{(1)}_u(x)$ satisfies the equation 
\begin{align}
\label{fokkerh1}
\frac{\sigma ^2}{2} \frac{\partial^2g^{(1)}_{u}}{\partial x^2} &- \alpha (x-m)
\frac{\partial  g^{(1)}_{u}}{\partial x} + \alpha \nu(u)  g^{(1)}_{u}(x) \\  
&= \mu\left(\frac{1}{u}-1\right) \left(g^{(1)}_{0}(x) -x(g_0^{(0)}(x) -g_u^{(0)}(x))\right)
\notag \\
&= \mu\left(\frac{1}{u}-1\right) \left(g^{(1)}_{0}(x) +  {x}\frac{\rho u c(g)}{(1-\rho u)} \right)\notag
\end{align}
where the constant $\nu(u)$ is given by
$$
\nu(u) = \frac{\mu (1-u)(1-\rho u)}{\alpha u}.
$$

In a first step, we search for a particular solution to the ordinary differential equation
$$
\frac{\sigma ^2}{2} \frac{\partial^2\xi_{u}}{\partial x^2} - \alpha (x-m) \frac{\partial  \xi_{u}}{\partial x} + \alpha \nu(u)  \xi_{u}(x) =  {x}\frac{\rho \mu (1-u) c(g)}{(1-\rho u)} 
$$
of the form
$$
\xi_u(x) = a(u)+b(u)x.
$$
Straightforward manipulations show that
$$
b(u) =\frac{\rho \mu (1-u)  (1-\rho)}{\alpha (\nu(u)-1)c(g)} \quad \mbox{and}\quad a(u) = -\frac{m}{\nu(u)}b(u). 
$$

Noting that $\xi_0(x) \equiv 0$, it follows that if we write $g_u^{(1)}(x) =
\xi_u(x)+\psi_u(x)$, then the function $\psi_u(x)$ is solution to the equation 
\begin{equation}
\label{fokkerpsi}
\frac{\sigma ^2}{2} \frac{\partial^2\psi_{u}}{\partial x^2} - \alpha (x-m) \frac{\partial  \psi_{u}}{\partial x} + \alpha \nu(u)  \psi_{u}(x) = \mu\left(\frac{1}{u}-1\right) \psi_0(x) .
\end{equation}
By using the domination property of Lemma~\ref{domg}, we can determine
the form of the function $\psi_0(x)$. 

\begin{lemma}
\label{expressh10}
The function $\psi_0(x)$ is given by
$$
\psi_0(x) = c_0 + c_1 \frac{\sqrt{\alpha}(x-m)}{\sigma}
$$
for some constants $c_0$ and $c_1$.
\end{lemma}

\begin{proof}
By introducing the function $k_u(x)$ defined by
\begin{equation}
\label{defku}
k_u(x) = \exp\left(-\frac{\alpha(x-m)^2}{2\sigma^2}\right) \psi_u(x)
\end{equation}
and then  the change of variable
\begin{equation}
\label{varchange}
z = \frac{\sqrt{\alpha}(x-m)}{\sigma},
\end{equation}
Equation~\eqref{fokkerpsi} becomes
\begin{equation}
\label{fokkerk0p}
\frac{\partial^2k_{u}}{\partial z^2} +(2\nu(u)+1 -z^2)  k_{u}(z) = \frac{2\mu}{\alpha}
\left(\frac{1}{u}-1\right)k_0(z). 
\end{equation}
The homogeneous equation reads
$$
\frac{\partial^2k_{u}}{\partial z^2}+(2\nu(u)+1 - z^2)  k_{u}=0,
$$
which solutions are parabolic cylinder functions (see Lebedev~\cite{Lebedev} for details). Two
independent solutions $v_1(u;z)$ and $v_2(u;z)$ of this homogeneous equation are given in
terms of Hermite functions as 
\begin{equation}
\label{defv1v2}
v_1(u;z) = e^{-{z^2}/{2}}H_{\nu(u)}(z) \quad \mbox{and} \quad v_2(u;z) = e^{{z^2}/{2}} H_{-\nu(u)-1}(iz).
\end{equation}
The Wronskian $\mathcal{W}$ of these two functions is given by
$$
\mathcal{W}(z) = e^{-{(\nu+1)\pi i}/{2}}.
$$

By using the method of variation of parameters, the solution to Equation~\eqref{fokkerk0p} is given by
\begin{multline*}
k_u(z) = \gamma_1(u)v_1(u;z) + \gamma_2(u) v_2(u;z) \\ 
-\frac{2\mu}{\alpha}\left(\frac{1}{u}-1\right) e^{{(\nu+1)\pi i}/{2}} \int_0^z \left[ v_1(u;y)v_2(u;z) -v_1(u;z)v_2(u;y)\right] k_0(y)\,dy,
\end{multline*}
where $\gamma_1(u)$ and $\gamma_2(u)$ are constants, which depend upon $u$.

The function $\psi_u(x)$ enjoys the same domination property as function $g_u^{(1)}(x)$,
given by Lemma~\ref{domg}. Hence, for $N>1$ 
\begin{equation}
\label{domk}
\lim_{z \to \pm  \infty} \frac{1}{z^N} e^{{z^2}/{2}} k_u(z) = 0.
\end{equation}

From Lebedev \cite{Lebedev}, we have the following asymptotic estimates
\begin{equation}
\label{estimeHermite}
H_\nu(z) \sim (2z)^\nu \left[\sum_{k=0}^n \frac{(-1)^k}{k!} (-\nu)_{2k}(2z)^{-2k} + O(|z|^{-2n-2})  \right]
\end{equation}
when $|z|\to \infty$ and $|\arg z|\leq {3\pi}/{4}-\delta$ for some $\delta>0$. Moreover, when $z \to -\infty$ 
$$
H_\nu(z) \sim \left\{ \begin{array}{l}  \displaystyle \frac{\sqrt{\pi}}{\Gamma(-\nu)}|z|^{-\nu-1}
  e^{z^2}\left[\sum_{k=0}^n \frac{1}{k!} (\nu+1)_{2k}(2z)^{-2k} + O(|z|^{-2n-2})  \right]      , \quad \nu \notin \N \\ \\ (2z)^\nu, \quad \nu \in \N . \end{array} \right. 
$$
The above asymptotic estimates and Lemma~\eqref{domg} imply that for $u\in(0,1)$ such that $\nu(u)\in \N$ with $\nu>1$,
we have 
\begin{eqnarray*}
\gamma_1(u) &=& - \frac{2\mu}{\alpha}\left(\frac{1}{u}-1\right) e^{{(\nu+1)\pi i}/{2}} \int_0^\infty v_2(u;y) k_0(y)\,dy \\
&=& - \frac{2\mu}{\alpha}\left(\frac{1}{u}-1\right) e^{{(\nu+1)\pi i}/{2}} \int_0^{-\infty} v_2(u;y) k_0(y) \,dy
\end{eqnarray*}
and 
\begin{eqnarray*}
\gamma_2(u) &=&\frac{2\mu}{\alpha}\left(\frac{1}{u}-1\right) e^{{(\nu+1)\pi i}/{2}}  \int_0^\infty v_1(u;y)k_0(y) \,dy \\
&=& \frac{2\mu}{\alpha}\left(\frac{1}{u}-1\right)  e^{{(\nu+1)\pi i}/{2}} \int_0^{-\infty} v_1(u;y)k_0(y) \,dy .
\end{eqnarray*}

The latter equation implies that for all $n>1$
\begin{equation}
\label{ortho}
 \int_{-\infty}^\infty e^{-{y^2}/{2}}k_0(y)H_n(y)\, dy =0,
\end{equation}
where $H_n(x)$ is the $n$th Hermite polynomial. By Property~\eqref{domk}, the
function $y \to \exp({y^2}/{2})k_0(y)$ is in $L^2(\R,\exp(-y^2)\,dy)$.
Since Hermite polynomials form an orthogonal basis in this Hilbert space,
Equation~\eqref{ortho} entails that the function $y \to \exp({y^2}/{2})k_0(y)$ is
orthogonal to all Hermite polynomials $H_n$ with $n>1$ and then that this function belongs
to the vector space spanned by $H_0$ and $H_1$. Hence, function $k_0(z)$ should be of the
form 
$$
k_0(z) = (c_0 + c_1z)  e^{-{z^2}/{2}}
$$
for some constants $c_0$ and $c_1$ and the result follows.
\end{proof}

By using the above lemma, we are now able to establish the expression of $g_u^{(1)}(x)$.

\begin{proposition}
\label{expressg1ux}
The function $g^{(1)}_u(x)$ is given by 
\begin{multline}
\label{expressh1u}
g_u^{(1)} (x) =  \frac{c(g)(u_1-1)(\tilde{u}_1-\rho  u)(u-1)}{u_1\tilde{u}_1(1-\rho)(u-\tilde{u}_1)(1-\rho u_1)(1-\rho u)^2}m  \\ +  \frac{c(g)(1-u)}{(u-\tilde{u}_1)(1-\rho u_1)(1-\rho u)}x,
\end{multline}
where $u_1$ and  $\tilde{u}_1$ are the two real solutions to the quadratic equation
$$\rho u^2-\left(1+\rho+\frac{\alpha}{\mu}\right)u+1=0$$
with $0<u_1<1<\tilde{u}_1$ 
\end{proposition}

\begin{proof}
By taking into account Lemma~\ref{expressh10}, the  function $K_u(z)$ defined by 
$$
K_u(x) = g_u^{(1)}(x) \exp\left(- \frac{\alpha(x-m)^2}{2\sigma^2}\right)
$$
and the  change of variable~\eqref{varchange}, satisfies the equation
\begin{multline}
\label{fokkerku}
\frac{\partial^2K_{u}}{\partial z^2} +(2\nu(u)+1 -z^2)  K_{u}(z) \\ = \frac{2\mu}{\alpha}
\left(\frac{1}{u}-1\right) \left(c_0+c_1z +\frac{\rho u c(g)}{1-\rho u}
\left(\frac{\sigma z}{\sqrt{\alpha}}+m  \right) \right)e^{-{z^2}/{2}}. 
\end{multline}
We search for a particular solution of the form
$$
K_u(z) = \left (a(u)+b(u)z \right)e^{-{z^2}/{2}}.
$$
Straightforward computations yield
\begin{align*}
a(u)  &=\frac{1}{(1-\rho u)}\left(c_0 + \frac{\rho u c(g)}{1-\rho u}m \right), \\
b(u) &= \frac{(1-u)}{\rho(u-u_1)(u-\tilde{u}_1)}\left(c_1 +\frac{\rho\sigma  u c(g)}{\sqrt{\alpha}(1-\rho u)}\right).
\end{align*}
It follows that the general solution to the above equation can be written as
\begin{equation}
\label{formKuz}
K_u(z) = \left (a(u)+b(u)z \right)e^{-{z^2}/{2}} + \gamma_1(u) v_1(u;z) + \gamma_2(u) v_2(u;z),
\end{equation}
where the functions $v_1$ and $v_2$ are defined by Equation~\eqref{defv1v2} and the constants $\gamma_1(u)$ and $\gamma_2(u)$ depend  upon $u$. 

By differentiating once Equation~\eqref{formKuz} with respect to $z$ and using the fact
that the Wronskian of the functions $v_1(u;z)$ and $v_2(u;z)$ is $\exp[(\nu(u)+1)\pi i/2]$,
we can easily express $\gamma_1(u)$ and $\gamma_2(u)$ by means of $K_u(z)$, $a(u)$, and
$b(u)$. This shows that $\gamma_1(u)$ and $\gamma_2(u)$ are analytic in the open unit disk
deprived of the points 0 and $u_1$. From the asymptotic properties satisfied by the
functions $v_1$ and $v_2$, we know that $\gamma_1(u) = 0$ and $\gamma_2(u)=0$ for $u$ such
that $\nu(u)>1$. It follows that $\gamma_1(u) \equiv \gamma_2(u) \equiv 0$ for $|u|<1$.  

By using the fact that $g_u^{(1)} (x)$ has to be analytic in variable $u$ in the unit disk, we necessarily have 
$$
c_1 = -\frac{\rho \sigma u_1 c(g)}{\sqrt{\alpha}(1-\rho u_1)}
$$
and then,
$$
b(u)=\frac{\sigma c(g)(1-u)}{ \sqrt{\alpha} (u-\tilde{u}_1)(1-\rho u_1)(1-\rho u)}.
$$

Moreover, since $g^{(1)}_1(x) \equiv 0$, we have 
$$
c_0 = -\frac{\rho c(g)m}{1-\rho}
$$ 
and then,
\begin{equation}
\label{defau}
a(u) = \frac{\rho(u-1)c(g)}{(1-\rho) (1-\rho u)^2}m.
\end{equation}
By using the expressions of $a(u)$ and $b(u)$, the result follows.
\end{proof}

\subsubsection{Higher order terms}

We assume that $g^{(i)}_u(x)$ can be expressed as
\begin{equation}
\label{formgnux}
g^{(i)}_u(x)  = \sum_{j=0}^i c_{i,j}(u) h_j(x),
\end{equation}
where the function $h_j$ is defined by Equation~\eqref{defhj} and the
coefficients $c_{i,j}$ are analytic functions in variable $u$. \cbstart This assumption will be justified a posteriori. \cbend From  previous sections,
this representation is valid for $i=0,1$. If it is valid for $i-1$, then the function
$g_u^{(i)}(x)$, $i \geq 1$, satisfies the equation 
\begin{multline}
\label{fokkerhn}
\frac{\sigma ^2}{2} \frac{\partial^2g^{(i)}_{u}}{\partial x^2} - \alpha (x-m) \frac{\partial  g^{(i)}_{u}}{\partial x} + \alpha \nu(u)  g^{(i)}_{u}(x) \\ 
= \mu\left(\frac{1}{u}-1\right) \left(g^{(i)}_{0}(x) -x(g_0^{(i-1)}(x) -g_u^{(i-1)}(x))\right) .
\end{multline}

First note that by using the recurrence relation~\eqref{recHermite}
 satisfied by Hermite polynomials, it is easily checked that
$$
x(g_u^{(i-1)}(x) -g_0^{(i-1)}(x)) = \sum_{j=0}^i d_{i,j}(u)h_j(x),
$$
where 
$$
d_{i,i}(u) = \frac{\sigma\sqrt{i}}{2\sqrt{\alpha}} (c_{i-1,i-1}(u) - c_{i-1,i-1}(0)),
$$
and for $0\leq j \leq i-1$,
\begin{multline*}
d_{j,i}(u) = \frac{\sigma\sqrt{j}}{2\sqrt{\alpha}}  (c_{i-1,j-1}(u) -c_{i-1,j-1}(0)) + m(c_{i-1,j}(u) -c_{i-1,j}(0))   \\  
 +  \frac{\sqrt{j+1} \sigma}{\sqrt{\alpha}}  (c_{i-1,j+1}(u) -c_{i-1,j+1}(0)) .
\end{multline*}
By using the above notation, we have the following result.

\begin{proposition}
\label{recgu}
The coefficients $c_{i,j}$ appearing in the representation~\eqref{formgnux} of $g_u^{(i)}(x)$ are recursively defined as follows: we have
$$
c_{0,0}(u) =  \frac{1-\rho}{1-\rho u},
$$
and for $i \geq 1$,
\begin{eqnarray*}
c_{i,0}(u) &=& \frac{d_{i,0}(u)-d_{i,0}(1)}{1-\rho u},\\
c_{i,j}(u) &=& \frac{\mu}{\alpha} \left(\frac{1}{u}-1\right) \frac{d_{i,j}(u)-   d_{i,j}(u_j) }{\nu(u)-j}\quad 1\leq j\leq i,
\end{eqnarray*}
where for $j \geq 1$, $u_j$ and $\tilde{u}_j$ are the two real solutions to the quadratic
equation $\nu(u)=j$, i.e.
$$
\rho u^2 -\left(1+\rho+\frac{j\alpha}{\mu}  \right)u+1=0
$$
with $0<u_j<1<\tilde{u}_j$. 
\end{proposition}

\begin{proof} As in the previous section, we first search for a solution to the equation
\begin{multline*}
\frac{\sigma ^2}{2} \frac{\partial^2\xi^{(i)}_{u}}{\partial x^2} - \alpha (x-m) \frac{\partial  \xi^{(i)}_{u}}{\partial x} + \alpha \nu(u)  \xi^{(i)}_{u}(x) \\ 
= \mu\left(\frac{1}{u}-1\right) x(g_u^{(i-1)}(x) -g_0^{(i-1)}(x)) .
\end{multline*}
Assuming that the function $\xi_u^{(i)}(x)$ is of the form
$$
\xi_u^{(i)}(x) = \sum_{k=0}^i\delta_{i,j}(u)h_j(x),
$$
we have, by using the fact that the functions $h_j(x)$ are eigenfunctions of the
operator $H$ associated with the eigenvalues $-\alpha j$ and that these functions are
linearly independent, for $j = 0, \ldots,i$, 
$$
\delta_{i,j} =  \frac{\mu}{\alpha} \left(\frac{1}{u}-1\right) \frac{d_{j,k}(u)}{\nu(u)-j}.
$$
It is easily checked that $\xi_0^{(i)}(x) \equiv 0$. We can then decompose $g_u^{(i)}(x)$ as
as 
$$g_u^{(i)}(x) = \psi_u^{(i)}(x) + \xi_u^{(i)}(x) ,$$ where the function $ \psi_u^{(i)}(x) $ is solution to the equation
$$
\frac{\sigma ^2}{2} \frac{\partial^2\psi^{(i)}_{u}}{\partial x^2} - \alpha (x-m) \frac{\partial  \psi^{(i)}_{u}}{\partial x} + \nu(u)  \psi^{(i)}_{u}(x) 
= \mu\left(\frac{1}{u}-1\right) \psi^{(i)}_0(x).
$$
By using the same arguments as in the proof of Lemma~\ref{expressh10}, we can easily show that  $\psi^{(i)}_0(x)$ has the form
$$
\psi^{(i)}_0(x) = \sum_{j=0}^i c_j h_j(x),
$$
where the coefficients $c_j \in \C$ for $j=0,...,i$. It follows that the function $g_u^{(i)}(x)$ is solution to the ordinary differential equation
$$
\frac{\sigma ^2}{2} \frac{\partial^2g^{(i)}_{u}}{\partial x^2} - \alpha (x-m) \frac{\partial  g^{(i)}_{u}}{\partial x} + \alpha \nu(u)  g^{(i)}_{u}(x) 
= \mu\left(\frac{1}{u}-1\right) \sum_{j=0}^i (c_j+d_{i,j}(u))h_j(x) .
$$

By using the same arguments as in the proof of Proposition~\ref{expressg1ux}, we come up with the conclusion that $g_u^{(i)}(x)$ is of the form~\eqref{formgnux}
with the coefficients $c_{i,j}(u)$ given by
$$
c_{i,j}(u) = \frac{\mu}{\alpha} \left(\frac{1}{u}-1\right) \frac{c_j + d_{i,j}(u)}{\nu(u)-j}.
$$
Since the function $g_u^{(i)}(x)$ has to be analytic in the open unit disk, we have for $j \geq 1$
$$
c_j = - d_{i,j}(u_j)
$$
In addition, since $g_1^{(i)}(x) \equiv 0$, we have $c_0 =  - d_{i,j}(1)$. 
\end{proof}

\cbstart
The normalizing constant $c(g)$ is chosen such that 
$$
\int_{-\infty}^\infty g_1(x)n(x)dx= 1.
$$
From the above analysis, we see that $g_1^{(i)}(x)\equiv 0$ for $i \geq 1$ so that $c(g)=1-\rho$.
\cbend

\subsubsection{Radius of convergence} In this section, we examine under which conditions the
expansion~\eqref{expandg} defines an element of $L^2(\R,n,\ell^2(\rho))$. In a first step, note that
as a consequence of Proposition~\ref{recgu}, the function  $g_u^{(i)}(x)$ can be written
as   
$$
g_u^{(i)}(x) = x \Theta \left(g_u^{(i-1)}(x)\right) = \Theta \left(x g_u^{(i-1)}(x)\right) 
$$
where the operator $\Theta$ is defined in $L^2(\R,n;\ell^2(1/\rho))$ as follows: for an element $f \in L^2(\R,n;\ell^2(1/\rho))$ represented as 
$$
f_u(x) \stackrel{def}{=} \sum_{j=0}^\infty c_j(u) h_j(x),
$$ 
the element $F = \Theta f$ is defined by 
$$
F_u(x) = \sum_{j=0}^\infty \mu\left( \frac{1}{u}-1 \right) \frac{c_j(u)-c_j(u_j)}{\nu(u)-j} h_j(x),
$$
where we set $u_0=1$ and $\tilde{u}_0 = 1/\rho$. 

It is easily checked that for $j\geq 1$, $0<u_j<1<1/\sqrt{\rho}<\tilde{u}_j$. Moreover,
the function $c_j(u)$ appearing in the expression of $f_u$ is analytic in the disk $D_\rho
= \{z : |z|< 1/\sqrt{\rho}\}$ and continuous in the closed disk $\overline{D}_\rho = \{z :
|z| \leq 1/\sqrt{\rho}\}$ for $j\geq 0$. Similarly, for all $j \geq 0$, the function  
$$
u \to \frac{\mu}{\alpha}\left( \frac{1}{u}-1 \right) \frac{c_j(u)-c_j(u_j)}{\nu(u)-j} 
$$
is analytic in $D_\rho$ and continuous in $\overline{D}_\rho$. With the above notation, we can state the main result of this section.

\begin{proposition}
The operator $\Theta$ is bounded and if $\varepsilon <1/(m \|\Theta\|)$, where $\|\Theta\|$ denotes the norm of $\Theta$, then the sequence  defined by Equation~\eqref{expandg} (or equivalently by Equation~\eqref{expandgu}) is in $L^2(\R,n;\ell^2(\rho))$.
\end{proposition}

\begin{proof}
Let $f \in L^2(\R,n;\ell^2(1/\rho))$ be defined by the function 
$$
f_u(x) = \sum_{j=0}^\infty c_j(u) h_j(x).
$$
For $(c_j) \in \ell^2(1/\rho)$ associated with the generating function
$$
c(u) = \sum_{j=0}^\infty c_ju^j ,
$$
we have
$$
\|c\|^2_{1/\rho} = \frac{1}{2\pi} \int_0^{2\pi}\left| c\left(\frac{1}{\sqrt{\rho}}e^{i\theta}\right)\right|^2d\theta.
$$
Let us moreover define the sequence $(\tilde{c}_j)$ associated with the generating function
$$
\tilde{c}(u) = \frac{\mu}{\alpha} \left( \frac{1}{u}-1 \right) \frac{c(u)-c(u_j)}{\nu(u)-j}.
$$
Assume first that $j \geq 1$, then
$$
\tilde{c}(u) = \frac{1}{\rho}(1-u)\frac{1}{u-\tilde{u}_j}\frac{c(u)-c({u}_j)}{u-u_j}.
$$
and then
$$
\|\tilde{c}\|^2_{1/\rho} \leq \frac{1}{\rho^2}\left(1+ \frac{1}{\sqrt{\rho}}\right)^2\frac{1}{(\tilde{u}_j-1/\sqrt{\rho})^2}\frac{1}{2\pi}\int_0^{2\pi}\left|
\frac{c(e^{i\theta}/\sqrt{\rho})-c({u}_j)}{e^{i\theta}/\sqrt{\rho}-u_j}
\right|^2d\theta . 
$$
Simple manipulations show that
%\marg{Détail ?}
$$
\frac{1}{2\pi}\int_0^{2\pi}\left|
\frac{c(e^{i\theta}/\sqrt{\rho})-c({u}_j)}{e^{i\theta}/\sqrt{\rho}-u_j}
\right|^2d\theta \leq \|c\|^2_\rho\frac{1}{(1/\sqrt{\rho}-u_j)^2}\left(
1+\sqrt{\frac{1}{1-\rho u_j^2}} \right)^2. 
$$
It follows that $\|\tilde{c}\|_{1/\rho} \leq \kappa_j \|c\|_{1/\rho}$, where
\begin{align*}
\kappa_j &=  \frac{1}{\rho}\left(1+ \frac{1}{\sqrt{\rho}}\right)\frac{1}{(\tilde{u}_j-1/\sqrt{\rho})(1/\sqrt{\rho}-u_j)}\left( 1+\sqrt{\frac{1}{1-\rho u_j^2}} \right) \\
&= \frac{1+\sqrt{\rho}}{(1-\sqrt{\rho})^2 + \frac{n\alpha}{\mu}} \left( 1+\sqrt{\frac{1}{1-\rho u_j^2}} \right) .
\end{align*}
It is easily checked that the sequence $(\kappa_j)$ for $ n \geq 1$ is decreasing.

When $j=0$, we define
$$
\tilde{c}(u) = \frac{\mu}{\alpha} \left( \frac{1}{u}-1 \right) \frac{c(u)-c(1)}{\nu(u)} =  \frac{c(u)-c(1)}{1-\rho u}.
$$
It is then easily checked that $\|\tilde{c}\|_{1/\rho} \leq \kappa_0  \|c\|_{1/\rho}$, where
$$
\kappa_0 =  \frac{1}{1-\sqrt{\rho}}\left( 1+\sqrt{\frac{1}{1- \rho}} \right).
$$

Define $\kappa = \max\{\kappa_0,\kappa_1\}$. The above computations show that for all
$f\in \ell^2(1/\rho)$, $\|\Theta f\| \leq \kappa \|f\|$. It follows that the operator
$\Theta$ is bounded; its norm is denoted by $\|\Theta\| \stackrel{\mathrm{def.}}{=} \inf\{c >0 : \forall f \in
\mathcal{H}, \; \|\Theta f\| \leq c \|f\|\}$. The above computations show that 
\begin{equation}
\label{normTheta}
\| \Theta\| \leq  \frac{1+\sqrt{\rho}}{(1-\sqrt{\rho})^2 }\left( 1+\sqrt{\frac{1}{1- \rho}} \right).
\end{equation}

From the above computations, we deduce that
$$
\|g^{(i)}\| \leq \|\Theta\|^i \|c^{(0)*i}\|
$$
where the sequence $c^{(0)*i}$ is associated with the function
$$
\frac{1-\rho}{1-\rho u} x^i.
$$

Straightforward computations show that 
\begin{equation}
\label{normc0*}
\|c^{(0)*i}\|^2 = \left(\frac{\sigma}{2\sqrt{\alpha}}  \right)^{2i} H_{2i}\left(\frac{\sqrt{\alpha}m}{\sigma}  \right),
\end{equation}
where $H_i(x)$ is the $i$th Hermite polynomial. Using the asymptotic estimate~\eqref{estimeHermite}, we have
$$
\|c^{(0)*i}\| \sim m^i
$$ 
when $i \to \infty$. It follows that $\|c^{(i)}\| \leq a_i$ with $a_i \sim (\|\Theta\|m)^i$ as $i$ tends to infinity. It follows that the sequence defined by the expansion~\eqref{expandgu} is convergent in $L^2(\R,n;\ell^2(\rho))$ if $\varepsilon \|\Theta\| m<1$.
\end{proof}

\cbstart
\subsection{Proof of Theorem~\ref{theo1}}

We assume that Conditions~\eqref{condeps} and \eqref{condeps2} are satisfied. By observing that $\mean\left(u^{L(t)}\right) = (P,U)$, we have for $u \leq 1$
$$
\left| \mean\left(u^{L(t)}\right) - (g,U) \right| \leq \| P-g\| \| U\| = \frac{1}{1-\rho u} \| P-g\| \leq \frac{1}{1-\rho u}  D(\varepsilon)
$$ 
where $U$ is the vector of $L^2(\R,n,\ell^2(\rho))$ with the $j$th component equal to $u^j$ and $D(\varepsilon)$ is defined by Equation~\eqref{defDeps}.

From the power series expansion of $g$, we have
$$
\left| (g,U) -(g^{(0)},U)-\varepsilon (g^{(1)},U) \right|   \leq \sum_{j=2}^\infty \varepsilon^j \| g^{(i)}\| \|U\| \leq \frac{1}{1-\rho u} \sum_{j=2}^\infty \kappa(j)  \varepsilon^j,
$$
with
$$
\kappa(j) =\\ \left(\frac{1+\sqrt{\rho}}{(1-\sqrt{\rho})^2 }\left( 1+\sqrt{\frac{1}{1- \rho}} \right)\right)^j \left(\frac{\sigma}{2\sqrt{\alpha}}  \right)^{2i} H_{2i}\left(\frac{\sqrt{\alpha}m}{\sigma}  \right),
$$
where we have used Equations~\eqref{normTheta} and \eqref{normc0*}. We clearly have
\begin{eqnarray*}
(g^{(0)},U) &=& \int_{-\infty}^\infty g_u^{(0)}(x) n(x) dx = \frac{1-\rho}{1-\rho u}\\
(g^{(1)},U) &=& \int_{-\infty}^\infty g_u^{(1)}(x) n(x) dx  = a(u) = -  \frac{\rho (1-u)}{(1-\rho u)^2}m,
\end{eqnarray*}
where $a(u)$ is defined by Equation~\eqref{defau}.  Theorem~\ref{theo1} then follows.

To complete the analysis, note that the generating function of $L_\varepsilon$, the stationary number of customers in an $M/M/1$ with input rate $\lambda$ and service rate $\mu(1-m\varepsilon)$,  is for $|u|<1$ and when $\varepsilon < (1-\rho)/m$ 
$$
\E\left( u^{L_\varepsilon} \right) -(g^{(0)},U)- (g^{(1)},U)\eps= \frac{\rho(1-u) m^2}{(1-\rho-m\eps)(1-\rho)^2}\eps^2
$$
Hence, by gathering these relations and by taking $u=1$,  we obtain an uniform bound
for the difference between $\mean\left(u^{L(t)}\right)$ and $\E\left( u^{L_\varepsilon} \right)$ for $u \in [0,1]$.   
\begin{multline}\label{Err}
\sup_{0\leq u\leq 1}\left| \mean\left(u^{L(t)}\right) - \E\left( u^{L_\varepsilon} \right) \right| \\ \leq 
E_B\stackrel{\text{def.}}{=}\frac{1}{1-\rho }  D(\varepsilon) + \frac{1}{1-\rho } \sum_{j=2}^\infty \kappa(j)
\varepsilon^j+\frac{2\rho m^2}{(1-\rho-m\eps)(1-\rho)^2}\eps^2
%%+ \sum_{j=2}^\infty \frac{2\rho }{(1-\rho u)^{n+1}} m^n\varepsilon^n. 
\end{multline}
Below are some numerical experiences on the role of $\eps$ and $\sigma$ on the bound $E_B$ for
$a=1/2$, $b= 1$ $m=1$, $\lambda=7$, $\alpha=1$ and $\mu=10$. It is reasonably low for
small values of $\eps$,  it seems to be quite sensitive on the values of the parameter
$\sigma$ as the figures below show.

\begin{figure}[ht]
\rotatebox{-90}{\scalebox{0.4}{\includegraphics{epsilon.ps}}}
\put(50,50){$\mathbf{\sigma=2}$}
\put(50,110){$\mathbf{\sigma=3}$}
\put(50,190){$\mathbf{\sigma=4}$}
\put(85,35){$\mathbf{\eps}$}
\put(-135,170){$\mathbf{E_B}$}
\vspace{-2cm}
\caption{The bound $\eps\to E_B$ of Relation~\eqref{Err} for $\sigma=2$, $3$, $4$.}
\end{figure}

\begin{figure}[ht]
\rotatebox{-90}{\scalebox{0.4}{\includegraphics{sigma.ps}}}
\put(80,45){$\mathbf{x=1}$}
\put(80,65){$\mathbf{x=2}$}
\put(80,90){$\mathbf{x=3}$}
\put(80,190){$\mathbf{x=5}$}
\put( 95,30){$\mathbf{\sigma}$}
\put(-125,170){$\mathbf{E_B}$}
\vspace{-1cm}
\caption{The bound $\sigma\to E_B$ of Relation~\eqref{Err} for $\eps=x\cdot 10^{-4}$ with $x=1,2,3,5$.}
\end{figure}
\cbend

\section{Concluding remarks}
\label{conclusion}

\cbstart The perturbation analysis performed in this paper has allowed us to prove the validity of the so-called reduced service rate approximation for the system considered under some specific conditions. Such an approximation is very important from a practical point of view because each type of traffic can be considered in isolation, the impact of unresponsive traffic on elastic traffic is only via the mean value.   \cbend 

The results presented in this paper have been obtained for a particular form
of the perturbation function $\phi(x)$. Of course, the same approach could be extended to
more complicated perturbation functions of the form $\Phi(x) = 1-\varepsilon p(x)$ for
some function $p(x)$. The key point consists of determining how the operator corresponding
to the multiplication by $p(x)$ acts on the basic functions $h_j(x)$ for $j \geq 0$ (defined by Equation~\eqref{defhj}). For computing
explicit expressions, however, the main difficulty is in solving the differential
equations satisfied by the coefficients of the expansion. When $p(x)$ is a polynomial, a
particular solution to the equations similar to Equations~\eqref{fokkerh1}
and~\eqref{fokkerhn} is obtained in the form of a polynomial times the function 
$\exp(-\alpha(x-m)^2/\sigma^2)$ and in that case, explicit computations can be  carried
out. 

The perturbation function $\phi(x)$ defined by Equation~\eqref{formphi} corresponds to the case when
unresponsive flows have a peak bit rate $\varepsilon$ much smaller than the transmission
capacity of the link. The results of this paper show that the reduced service rate
approximation yields in some conditions  accurate results for the performance of elastic  flows. 

\appendix

\section{Proof of Proposition~\ref{PropalPhi}}
\label{demPhi}

To prove Proposition~\ref{PropalPhi}, we proceed as for the proof of Proposition~\ref{Propalphi}. We first show that the operator $(H_1+A_1)^{-1}W_1$ is bounded, where $W_1$ is the restriction to $\ell_1^2(\rho)$ of the operator $W$.  (Note that the operator associated with the multiplication by $\Phi(x)-1$ is not bounded in $L^2(\R,n)$.) For this purpose we use the fact  that an element of $f = (f_j(x)) \in L^2(\R,n;\ell^2_1(\rho))$ can be decomposed as
$$
f = \sum_{j=1}^\infty \sum_{k=0}^\infty  c_{j,k}    h_k(x) e_j.
$$
and the squared norm is 
$$
\|f\|^2 =  \sum_{j=1}^\infty \sum_{k=0}^\infty  c^2_{j,k}\rho^j.
$$

By using the above decomposition, we have
$$
Bf =\sum_{j=1}^\infty \sum_{k=0}^\infty  (-\mu c_{j,k} +\lambda c_{j+1,k})   h_k(x) e_j
$$
By using the recurrence relation satisfied by Hermite polynomials
\begin{equation}
\label{hermiterel}
xH_k(x) = \frac{1}{2}H_{k+1}(x) +kH_{k-1}(x),
\end{equation}
we deduce that 
$$
x h_k(x) = \frac{\sigma}{\sqrt{2\alpha}}\left(\sqrt{\frac{k+1}{2}} h_{k+1}(x) +\sqrt{\frac{k}{2}}h_{k-1}(x)  \right) +mh_k(x).
$$
and then
$$
W_1f =\varepsilon \sum_{j=1}^\infty  \sum_{k=0}^\infty C_{j,k} h_k(x) e_j,
$$
where
\begin{multline*}
C_{j,k} = m\left(\mu c_{j,k}-\lambda c_{j+1,k}\right) + \\
 \frac{\sigma}{\sqrt{2\alpha}} \left(\ind_{\{k>0\}}\sqrt{\frac{k}{2}} \left(\mu c_{j,k-1}-\lambda c_{j+1,k-1}\right) +\sqrt{\frac{k+1}{2}}\left(\mu c_{j,k+1}-\lambda c_{j+1,k+1}\right ) \right).
\end{multline*}
From the above relations, we have
\begin{multline}
\label{decompotech}
(H_1+A_1)^{-1}W_1f = \\ \varepsilon \left( \sum_{j=1}^\infty  C_{j,0} (H_1+A_1)^{-1}   h_0(x) e_j + \sum_{j=1}^\infty \sum_{k=1}^\infty  C_{j,k} (H_1+A_1)^{-1}   h_k(x) e_j   \right) 
\end{multline}

For the first term in the right hand side of Equation~\eqref{decompotech}, we have
\begin{multline*}
\left\|\sum_{j=1}^\infty  C_{j,0} (H_1+A_1)^{-1}   h_0(x) e_j \right\| \leq \left\|(H_1+A_1)^{-1}\right\| \left\|\sum_{j=1}^\infty  C_{j,0} h_0(x) e_j\right\| \\ 
= \left\|(H_1+A_1)^{-1}\right\|  \sqrt{\sum_{j=1}^\infty  C_{j,0}^2 \rho^j}  \leq  \mu(1+\sqrt{\rho}) \left(m+\frac{\sigma}{2\sqrt{\alpha}}\right) \left\|(H_1+A_1)^{-1}\right\| \|f\|,
\end{multline*}
since
$$
\sum_{j=1}^\infty  C_{j,0}^2 \rho^j \leq \left(m \sqrt{\sum_{j=1}^\infty(\mu c_{j,0}-\lambda c_{j+1,0})^2\rho^j}   +\frac{\sigma}{2\sqrt{\alpha}}  \sqrt{\sum_{j=1}^\infty(\mu c_{j,1}-\lambda c_{j+1,1})^2\rho^j} \right)^2
$$
together with the inequalities 
\begin{multline*}
\sqrt{\sum_{j=1}^\infty(\mu c_{j,0}-\lambda c_{j+1,0})^2\rho^j} \leq \left(\mu\sqrt{\sum_{j=1}^\infty c_{j,0}^2\rho^j} +\sqrt{\lambda\mu}  \sqrt{\sum_{j=1}^\infty c_{j+1,0}^2\rho^{j+1}} \right) \\ \leq \mu(1+\sqrt{\rho})\|f\|,
\end{multline*}
and
$$
\sqrt{\sum_{j=1}^\infty(\mu c_{j,1}-\lambda c_{j+1,1})^2\rho^j}  \leq \mu(1+\sqrt{\rho})\|f\|.
$$

For the second term in the right hand side of Equation~\eqref{decompotech}, since the operator $H$ is invertible on the space $\mathrm{span}(h_k, k \geq 1)$, we have
$$
\sum_{j=1}^\infty \sum_{k=1}^\infty  C_{j,k} (H_1+A_1)^{-1}   h_k(x) e_j    = \sum_{j=1}^\infty \sum_{k=1}^\infty  C_{j,k} (\I+H_1^{-1}A_1)^{-1}  H_1^{-1} h_k(x) e_j
$$
and then, by using the fact that  $\|(\I+H_1^{-1}A_1)^{-1}\|\leq 1$, we obtain
$$
\left\| \sum_{j=1}^\infty \sum_{k=1}^\infty  C_{j,k} (H_1+A_1)^{-1}   h_k(x) e_j\right\| \leq   \sqrt{\sum_{j=1}^\infty \sum_{k=1}^\infty  \left(\frac{C_{j,k}}{k}\right)^2\rho^j}.
$$
From the inequality
\begin{multline*}
 \sqrt{\sum_{j=1}^\infty \sum_{k=1}^\infty  \left(\frac{C_{j,k}}{k}\right)^2\rho^j} \\ 
\leq m \sqrt{\sum_{j=1}^\infty \sum_{k=1}^\infty \frac{\left(\mu c_{j,k}-\lambda c_{j+1,k}\right)^2}{k^2}\rho^j}  +\frac{\sigma}{2\sqrt{\alpha}} \sqrt{\sum_{j=1}^\infty \sum_{k=1}^\infty  \frac{\left(\mu c_{j,k-1}-\lambda c_{j+1,k-1}\right)^2}{k}\rho^j} \\
+ \frac{\sigma}{2\sqrt{\alpha}} \sqrt{\sum_{j=1}^\infty \sum_{k=1}^\infty  \frac{(k+1)\left(\mu c_{j,k+1}-\lambda c_{j+1,k+1}\right )^2}{k^2}\rho^j}
\end{multline*}
we deduce that
$$
\sqrt{\sum_{j=1}^\infty \sum_{k=1}^\infty  \left(\frac{C_{j,k}}{k}\right)^2\rho^j} \leq \mu(1+\sqrt{\rho})\left(m +\frac{\sigma}{\sqrt{\alpha}}\right) \|f\|
$$

Finally, by using the fact that $\|(H_1+A_1)^{-1}\|\leq 1/(\mu(1-\sqrt{\rho})^2)$, we come up with the conclusion that
\begin{equation}
\label{normW1}
\|(H_1+A_1)^{-1}W_1f\| \leq  2  \varepsilon \frac{(1+\sqrt{\rho})}{(1-\sqrt{\rho})^2} \left(m+\frac{\sigma}{\sqrt{\alpha}}\right)\|f\|.
\end{equation}
The operator $(H_1+A_1)^{-1}W_1$ is hence bounded. Under Condition~\eqref{condeps2}, the norm of this operator is less than 1 and we can adapt word by word the proof of Proposition~\ref{Propalphi}.

\end{document}